\RequirePackage{xcolor}
\documentclass[journal,twoside,web]{IEEEtran}
\usepackage{etoolbox}
\makeatletter
\@ifundefined{color@begingroup}%
  {\let\color@begingroup\relax
   \let\color@endgroup\relax}{}%
\def\fix@ieeecolor@hbox#1{%
  \hbox{\color@begingroup#1\color@endgroup}}
\patchcmd\@makecaption{\hbox}{\fix@ieeecolor@hbox}{}{\FAILED}
\patchcmd\@makecaption{\hbox}{\fix@ieeecolor@hbox}{}{\FAILED}

\usepackage{tmi}
\usepackage{amsmath,amssymb,amsfonts}
\usepackage{algorithmic}
\usepackage{graphicx}
\usepackage{textcomp}
\usepackage[caption=false]{subfig}
\usepackage{bbm}
\usepackage{booktabs}
\usepackage{multirow}
\usepackage[ruled,vlined,linesnumbered,noresetcount]{algorithm2e}
\usepackage[colorinlistoftodos,prependcaption]{todonotes}
\usepackage{bm}
\usepackage{scalerel}
\usepackage{tikz}
\usetikzlibrary{svg.path}

\definecolor{orcidlogocol}{HTML}{A6CE39}
\tikzset{
  orcidlogo/.pic={
    \fill[orcidlogocol] svg{M256,128c0,70.7-57.3,128-128,128C57.3,256,0,198.7,0,128C0,57.3,57.3,0,128,0C198.7,0,256,57.3,256,128z};
    \fill[white] svg{M86.3,186.2H70.9V79.1h15.4v48.4V186.2z}
                 svg{M108.9,79.1h41.6c39.6,0,57,28.3,57,53.6c0,27.5-21.5,53.6-56.8,53.6h-41.8V79.1z M124.3,172.4h24.5c34.9,0,42.9-26.5,42.9-39.7c0-21.5-13.7-39.7-43.7-39.7h-23.7V172.4z}
                 svg{M88.7,56.8c0,5.5-4.5,10.1-10.1,10.1c-5.6,0-10.1-4.6-10.1-10.1c0-5.6,4.5-10.1,10.1-10.1C84.2,46.7,88.7,51.3,88.7,56.8z};
  }
}
\newcommand{\croppdf}[1]{\IfFileExists{#1-crop.pdf}{}{\immediate\write18{pdfcrop #1.pdf}}}

\newcommand\orcidicon[1]{\href{https://orcid.org/#1}{
\mbox{\scalerel*{
\begin{tikzpicture}[shift={(3 cm,5 cm)}, yscale=-1,transform shape]
\pic{orcidlogo};
\end{tikzpicture}
}
{A}
}
}}

\SetKwInput{KwInput}{Input}                % Set the Input
\SetKwInput{KwOutput}{Output}              % set the Output

\usepackage[normalem]{ulem}

\newcommand{\update}[1]{#1}
\newcommand{\updatetwo}[1]{#1}
\newcommand{\remove}[1]{}
\newcommand{\removetwo}[1]{}
\newcommand{\removecref}[1]{}
\newcommand{\removecreftwo}[1]{}
\newcommand{\notetwo}[1]{}
\newcommand{\note}[1]{}
\newcommand{\linelabel}[1]{}

\newcommand{\etal}{\textit{et~al.}}
\newcommand{\subbest}[1]{\textbf{#1}}
\newcommand{\unsubbest}[1]{\underline{#1}}

\newcounter{question}

\makeatletter
\let\NAT@parse\undefined
\makeatother
\usepackage[hypertexnames=false]{hyperref} 
\usepackage[capitalise,nameinlink,noabbrev]{cleveref}
\hypersetup{
 colorlinks=true,
 linkcolor=mblue,
 urlcolor=cyan,
 citecolor=black
}

\def\BibTeX{{\rm B\kern-.05em{\sc i\kern-.025em b}\kern-.08em
    T\kern-.1667em\lower.7ex\hbox{E}\kern-.125emX}}
\begin{document}

\Crefname{figure}{Suppl. Figure}{Suppl. Figures}
\Crefname{table}{Suppl. Table}{Suppl. Tables}
\Crefname{section}{Suppl. Section}{Suppl. Sections}

% Uncomment to view responses
% \input{_responsesR2}
% \input{_revision_summaryR2}
% \newpage

\twocolumn

\title{Clinically-Inspired Multi-Agent Transformers \\ for Disease Trajectory Forecasting \\ from Multimodal Data}
\author{Huy Hoang Nguyen\orcidicon{0000-0001-9537-1784},
        Matthew B.\ Blaschko\orcidicon{0000-0002-2640-181X},
        Simo Saarakkala\orcidicon{0000-0003-2850-5484},
        and Aleksei Tiulpin\orcidicon{0000-0002-7852-4141}
%\thanks{This work was supported in part by the strategic funding of the University of Oulu, and in part by Sigrid Juselius Foundation, Finland.}
\thanks{Huy Hoang Nguyen is with the Research Unit of Health Sciences and Technology, University of Oulu, Finland. E-mail: huy.nguyen@oulu.fi.}
\thanks{Matthew Blaschko is with Center for Processing Speech \& Images, KU Leuven,  Belgium. E-mail: matthew.blaschko@esat.kuleuven.be.}
\thanks{Simo Saarakkala is with the Research Unit of Health Sciences and  Technology, 
 University of Oulu, Finland and Department of Diagnostic Radiology, Oulu University Hospital, Finland. E-mail: simo.saarakkala@oulu.fi.}
\thanks{Aleksei Tiulpin is with the Research Unit of Health Sciences and Technology, 
 University of Oulu, Finland and Neurocenter Oulu, Oulu University Hospital. E-mail: aleksei.tiulpin@oulu.fi}
 % \thanks{Copyright (c) 2021 IEEE. Personal use of this material is permitted. However, permission to use this material for any other purposes must be obtained from the IEEE by sending a request to pubs-permissions@ieee.org.}
}

\maketitle

% \linenumbers
\setcounter{page}{1}
\setcounter{figure}{0}
\setcounter{table}{0}

\begin{abstract}
\label{sc:abstract}
Deep neural networks are often applied to medical images to automate the problem of medical diagnosis. However, a more clinically relevant question that practitioners usually face is how to predict the future trajectory of a disease. Current methods for prognosis or disease trajectory forecasting often require domain knowledge and are complicated to apply. In this paper, we formulate the prognosis prediction problem as a one-to-many prediction problem. Inspired by a clinical decision-making process with two agents -- a radiologist and a general practitioner -- we predict prognosis with two transformer-based components that share information with each other. The first transformer in this framework aims to analyze the imaging data, and the second one leverages its internal states as inputs, also fusing them with auxiliary clinical data. The temporal nature of the problem is modeled within the transformer states, allowing us to treat the forecasting problem as a multi-task classification, for which we propose a novel loss. We show the effectiveness of our approach in predicting the development of structural knee osteoarthritis changes and forecasting Alzheimer's disease clinical status directly from raw multi-modal data. The proposed method outperforms multiple state-of-the-art baselines with respect to performance and calibration, both of which are needed for real-world applications. An open-source implementation of our method is made publicly available at \url{https://github.com/Oulu-IMEDS/CLIMATv2}.

\end{abstract}


% \fi

\IEEEpeerreviewmaketitle

\section{Introduction}
\IEEEPARstart{R}{ecent} developments in Machine Learning (ML) suggest that it is soon to be tightly integrated into many fields, including  healthcare~\cite{desai2020comparison,boutet2021predicting}. One particular subfield of ML -- Deep Learning (DL) has advanced the most, as it opened the possibility to make predictions from high-dimensional data. In medicine, this impacted the field of radiology, in which highly trained human readers identify pathologies in medical images. The full clinical pipeline, however, aims to assess the condition of a patient as a whole, and eventually prescribe the most relevant treatment for a disease~\cite{cheerla2019deep,tran2021deep}. Using DL in this broad scope by integrating multimodal data has the potential to provide even further advances in medical applications.

\begin{figure}[t]
    \centering
    \croppdf{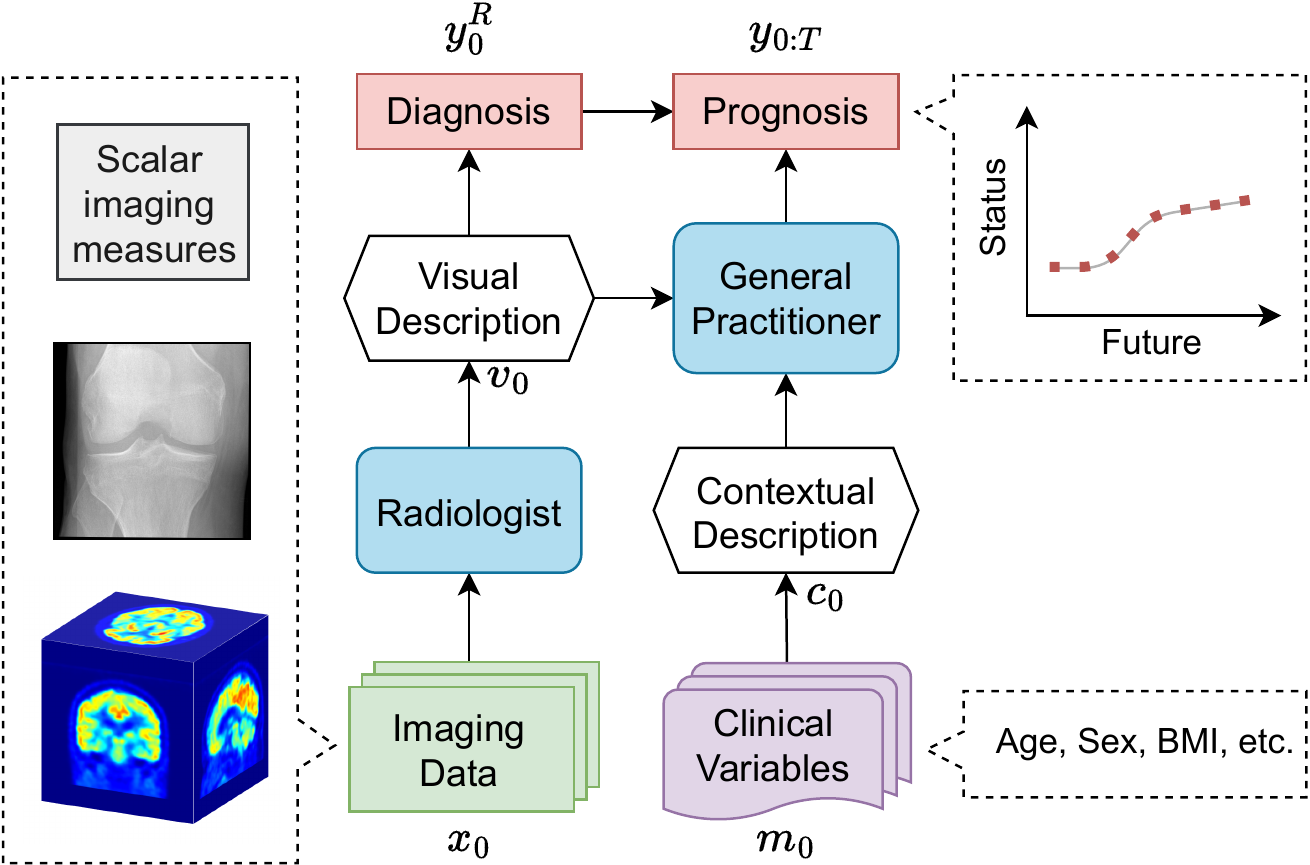}{}{\immediate\write18{pdfcrop figures/CLIMAT/CLIMAT_concept}}    
    \includegraphics[width=.48\textwidth]{figures/CLIMAT/CLIMAT_concept-crop.pdf}        
    \caption{\small The concept of CLIMATv2 was inspired by a multi-agent decision-making system with a radiologist and a general practitioner. All types of imaging data of disease are handled by the radiologist. The general practitioner then utilizes a report description produced by the radiologist and the context of clinical variables to forecast a future trajectory of the disease.}
    \label{fig:climat_concept}
\end{figure}

Clinical diagnosis is made by specialized treating physicians or general practitioners. These doctors are not radiologists and rather use the services of the latter in decision-making. One of the typical problems that such doctors face is to make a prognosis~\cite{milanez2020cancer,mei2020artificial}, which can be formalized as disease trajectory forecasting (DTF). This is an especially relevant task in degenerative disorders, often seen in musculoskeletal and nervous systems. This work studies DTF for knee osteoarthritis (OA) -- the most common musculoskeletal disorder~\cite{glyn2015osteoarthritis}, and Alzheimer's disease (AD) -- the leading cause of dementia~\cite{AD2021}.

\begin{figure}[t!]
    \centering
        \subfloat[BL - KL $0$]{\includegraphics[width=0.1\textwidth]{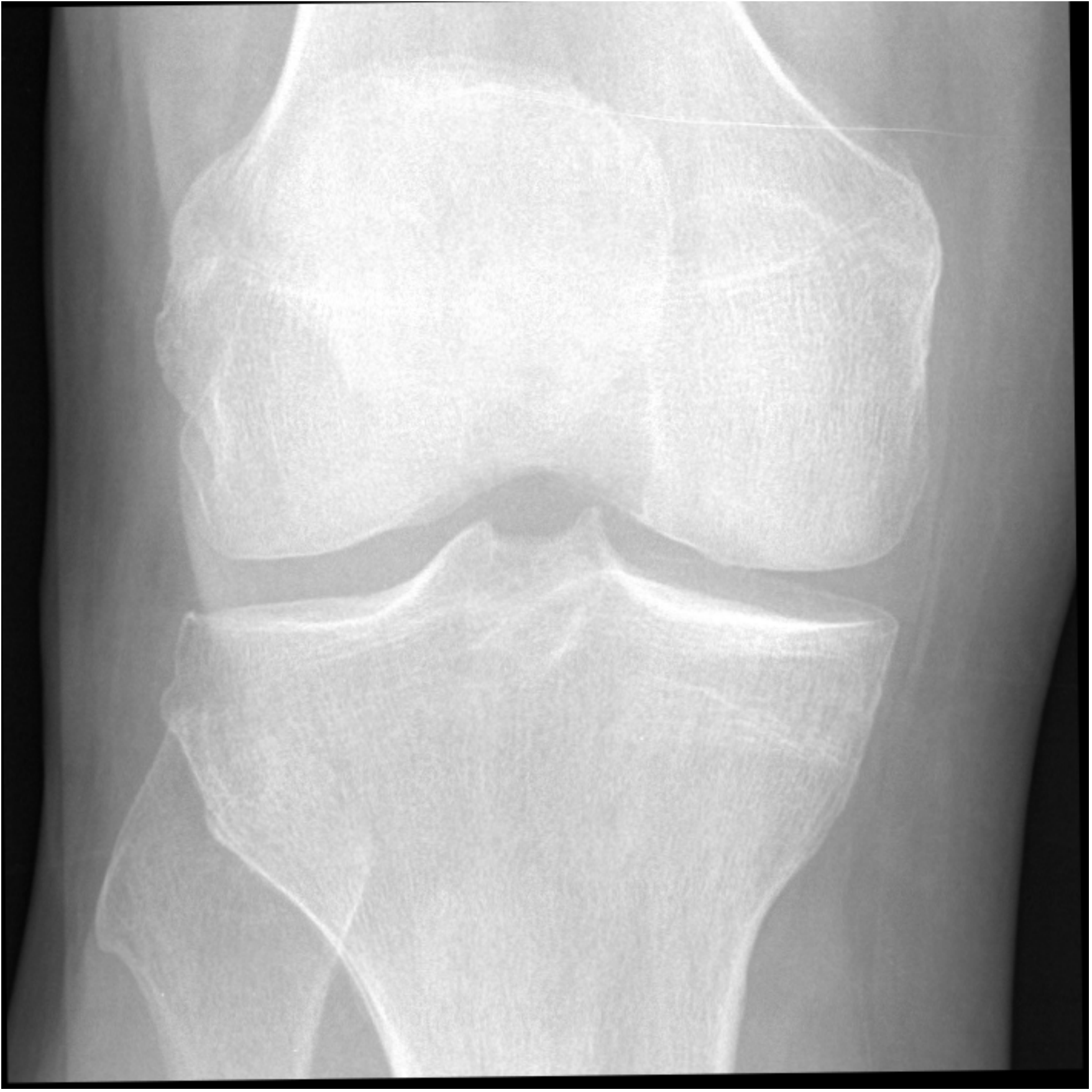}}\hfill%
        \subfloat[Y3 - KL $1$]{\includegraphics[width=0.1\textwidth]{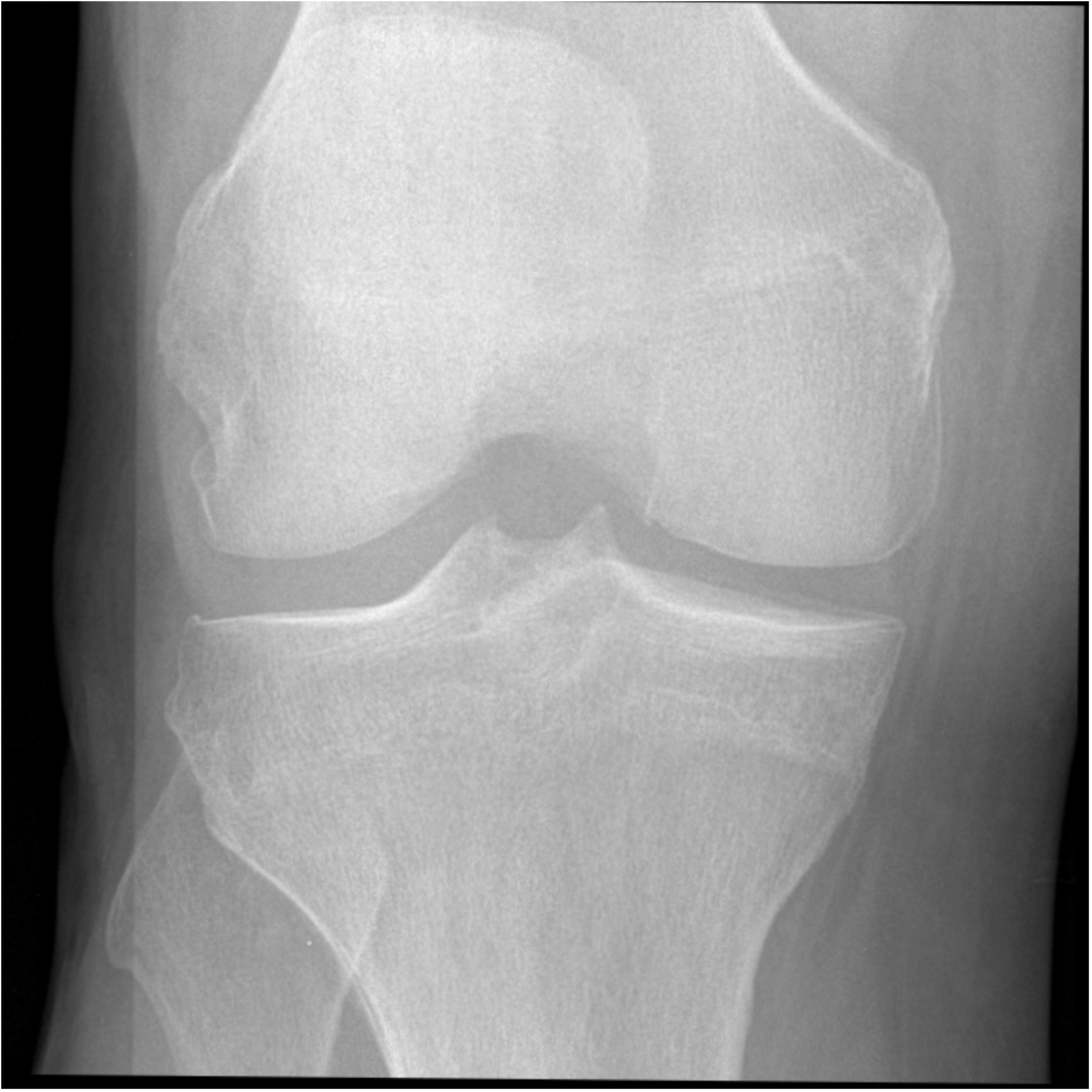}}\hfill%
        \subfloat[Y$6$ - KL $3$]{\includegraphics[width=0.1\textwidth]{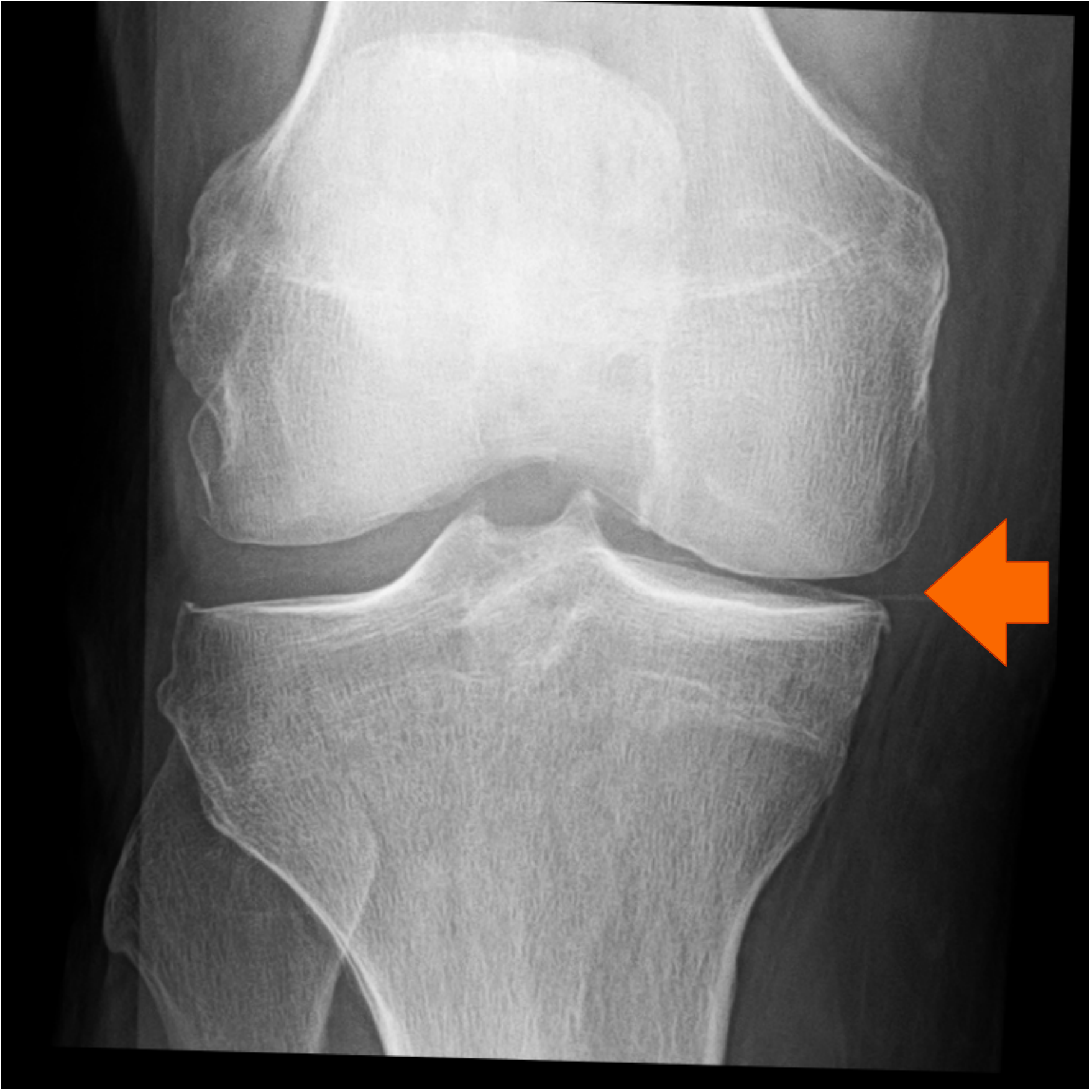}}\hfill%
        \subfloat[Y$8$ - TKR]{\includegraphics[width=0.1\textwidth]{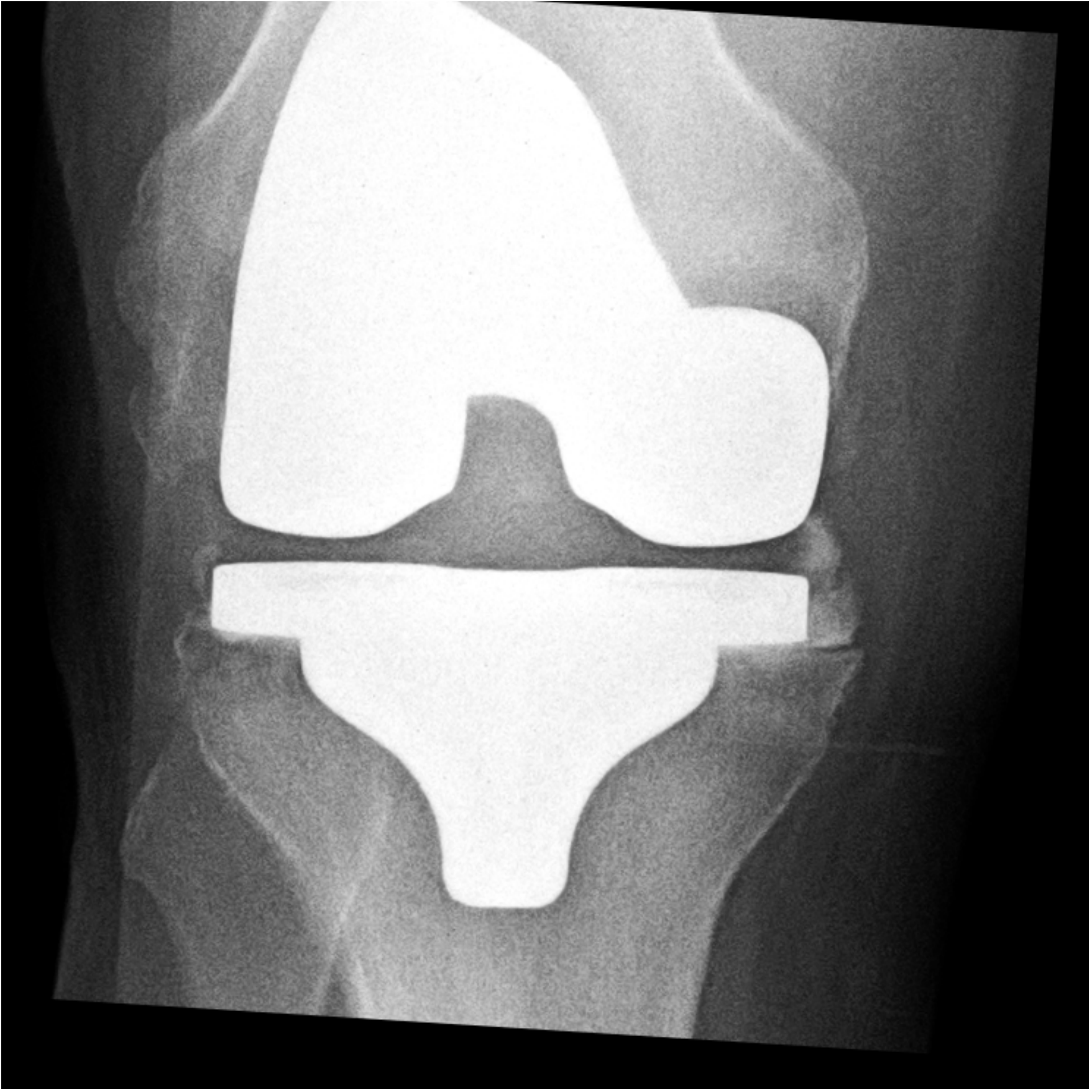}}\hfill%        
    \caption{\small Radiographs of a patient with knee OA progressed in $8$ years. The \remove{red}\update{orange} arrow indicates joint space narrowing. The disease progressed from Kellgren-Lawrence (KL) grade $0$ at the baseline (BL) to $3$ in $6$ years. At the $8$th year, the patient underwent a total knee replacement (TKR) surgery.}
    \label{fig:progression_sample}
\end{figure}

Among all the joints in the body, OA is mostly prevalent in the knee. Knee OA is characterized by the appearance of osteophytes, and the narrowing of joint space~\cite{heidari2011knee}, which in the clinical setting are usually imaged using X-ray (radiography)~\cite{lee2021imaging}. The disease severity is graded according to the Kellgren-Lawrence system~\cite{kellgren1957radiological} from 0 (no OA) to 4 (end-stage OA), or Osteoarthritis Research Society International (OARSI) atlas criteria~\cite{altman2007atlas}. Unfortunately, OA is progressive over time (see~\cref{fig:progression_sample}) and no cure has yet been developed for OA. However, diagnosing OA at an early stage may allow the slowing down of the disease, for example using behavioral interventions~\cite{rezucs2021pathogenesis}.

Individuals with AD have difficulties with reading, learning, and even performing daily activities. AD is fatally progressive and caused more than $120,000$ deaths in the United States in $2019$; however, no effective cure for it has been made available~\cite{AD2021}. The benefits of early AD diagnosis are similar to OA -- the progression of the disease can be delayed, and patients may be assigned relevant care in a timely manner~\cite{rasmussen2019alzheimer}.

In both of the aforementioned fields -- OA and AD, there is a lack of studies on prognosis prediction. From an ML perspective, a more conventional setup is to predict \textit{whether} the patient has the disease at present or a specific point of time in the future~\cite{tiulpin2019multimodal,widera2020multi,guan2020deep,leung2020prediction,tolpadi2020deep,jung2019unified,lu2018multimodal}. However, prognosis prediction aims to answer \textit{whether} and \textit{how} the disease would evolve over time. Furthermore, in a real-life situation, the treating physician makes the prognosis while interacting with a radiologist or other stakeholders who can provide information (e.g.\ blood tests or radiology reports) about the patient's condition. This also largely differentiates the diagnostic task from predicting a prognosis. 

In this paper, we present an extended version of our earlier work on automatic DTF~\cite{nguyen2022climat}, where we proposed a Clinically-Inspired Multi-Agent Transformers (CLIMAT) framework, aiming to mimic the interaction process between a general practitioner / treating physician\footnote{In the sequel we write general practitioner, which, however, does not restrict our modeling approach} and a radiologist. In our system, a radiologist module, consisting of a feature extractor (convolutional neural network; CNN) and a transformer, analyses the input imaging data and then provides an output state of the transformer representing a radiology report to the general practitioner -- corresponding module (purely transformer-based). The latter fuses this information with auxiliary patient data, and makes the prognosis prediction. We graphically illustrate the described idea in~\cref{fig:climat_concept}.

Compared to the conference version~\cite{nguyen2022climat}, we have enhanced our framework, such that the module corresponding to the general practitioner does not only perform prognosis, but is also encouraged to \removetwo{make diagnoses}\updatetwo{make diagnostic predictions} consistent with a radiologist module. \linelabel{ln:version_cmp}\updatetwo{The earlier version of CLIMAT relies on a simplifying assumption in relation to the independence between the diagnostic label task and non-imaging data.% In OA and AD~\cite{liu2023joint,bird2005genetic,li2022validation}, but also in other diseases~\cite{udler2019genetic,cole2020genetics,sule2020real,zhang2021insight}, it has been shown that as in clinical work, multimodal patient-level information is beneficial for diagnostic predictions.
} The \updatetwo{introduced} update helps the framework to expand out of the knee osteoarthritis domain, \updatetwo{and be more realistic, thereby allowing our method to be applied in fields} where diagnosis \removetwo{relies}\updatetwo{could rely} on both imaging and non-imaging data. Moreover, we equip the framework with a new loss -- Calibrated Loss based on Upper Bound (CLUB) -- that aims to maintain the performance while improving the calibration of the framework's predictions. Finally, we have also expanded the application of our framework to the case of AD.

To summarize, our contributions are the following:

\begin{enumerate}
    \item  We propose CLIMATv2, a clinically-inspired transformer-based framework that can learn to forecast disease severity from multi-modal data in an end-to-end manner. The main novelty of our approach is the incorporation of prior knowledge of the decision-making process into the model design.
    \item \linelabel{ln:2nd_contribution}We derive the CLUB loss, an upper bound on a temperature-scaled cross-entropy (TCE), and apply it to the DTF problem we have at hand.  Experimentally, we show that CLUB provides better calibration and yields similar or better balanced accuracy than the competitive baselines.
    \item From a clinical perspective, our results show the feasibility to perform fine-grained prognosis of knee OA and AD  directly from raw multi-modal 2D and 3D data.
\end{enumerate}

\section{Related Work}
\subsubsection{Knee osteoarthritis prognosis} 
The attention of the literature has gradually been shifting from diagnosing the current OA severity of a knee to predicting whether degenerative changes will happen within a specified time frame.
While some studies~\cite{tiulpin2019multimodal,widera2020multi,guan2020deep} aimed to predict whether knee OA progresses within a specified duration, others~\cite{leung2020prediction,tolpadi2020deep} tried to predict if a patient will undergo a total knee replacement (TKR) surgery at some point in the future. However, the common problem of the aforementioned studies is that the scope of knee OA progression is limited to a single period of time or outcome, which substantially differentiates our work from the prior art.  

\begin{figure}[t]
    \centering
    \hspace*{\fill}
        \subfloat[Axial]{\includegraphics[width=0.1\textwidth]{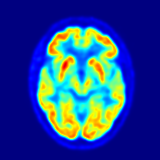}}\hfill
        \subfloat[Coronal]{\includegraphics[width=0.1\textwidth]{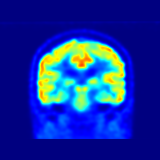}}\hfill
        \subfloat[Sagittal]{\includegraphics[width=0.1\textwidth]{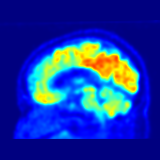}}
        \hspace*{\fill}
    \caption{\small The three projections of a 3D FDG-PET scan, which is converted to the jet colormap for demonstration purposes. The red regions are associated with high concentrations of the FDG radioactive tracer in the brain.}
    \label{fig:fdg_samples}
\end{figure}

\subsubsection{Alzheimer's disease prognosis}
Compared to the field of OA, a variety of approaches have been proposed to process longitudinal data in the AD field. Lu~\etal~\cite{lu2018multimodal} utilized a fully-connected network (FCN) to predict AD progression within a time frame of $3$ years from magnetic resonance imaging (MRI) and fluorodeoxyglucose positron emission tomography (FDG-PET) scans. Ghazi~\etal~\cite{ghazi2019training} and Jung~\etal~\cite{jung2019unified} used different long-short-term memory (LSTM)-based models to predict AD clinical statuses 
from scalar MRI biomarkers. Albright~\etal~\cite{albright2019forecasting} took into account various combinations of scalar measures and clinical variables to predict changes in AD statuses using FCNs and recurrent neural networks (RNN). In contrast to the prior art relying on either raw imaging data or scalar measures, our method enables learning from raw imaging scans, imaging-based measurements, and other scalar variables simultaneously. Additionally, whereas FCN and sequential networks were widely used in the literature, we propose to use a transformer-based framework to perform the AD clinical status prognosis task. Furthermore, we use FCN, two well-known sequential models -- gated recurrent unit (GRU) and LSTM -- as our reference approaches.

\begin{figure*}[t!]
    \centering
    \croppdf{figures/CLIMAT/CLIMAT_workflow_v1}
    \croppdf{figures/CLIMAT/CLIMAT_workflow_v2}    
    \hspace*{\fill}
    % \subfloat[CLIMATv1~\cite{nguyen2022climat}]{\includegraphics[height=0.45\textwidth]{figures/CLIMAT/CLIMAT_workflow_v1-crop.pdf}}
    \subfloat[CLIMATv1~\cite{nguyen2022climat}]{\includegraphics[height=0.45\textwidth]{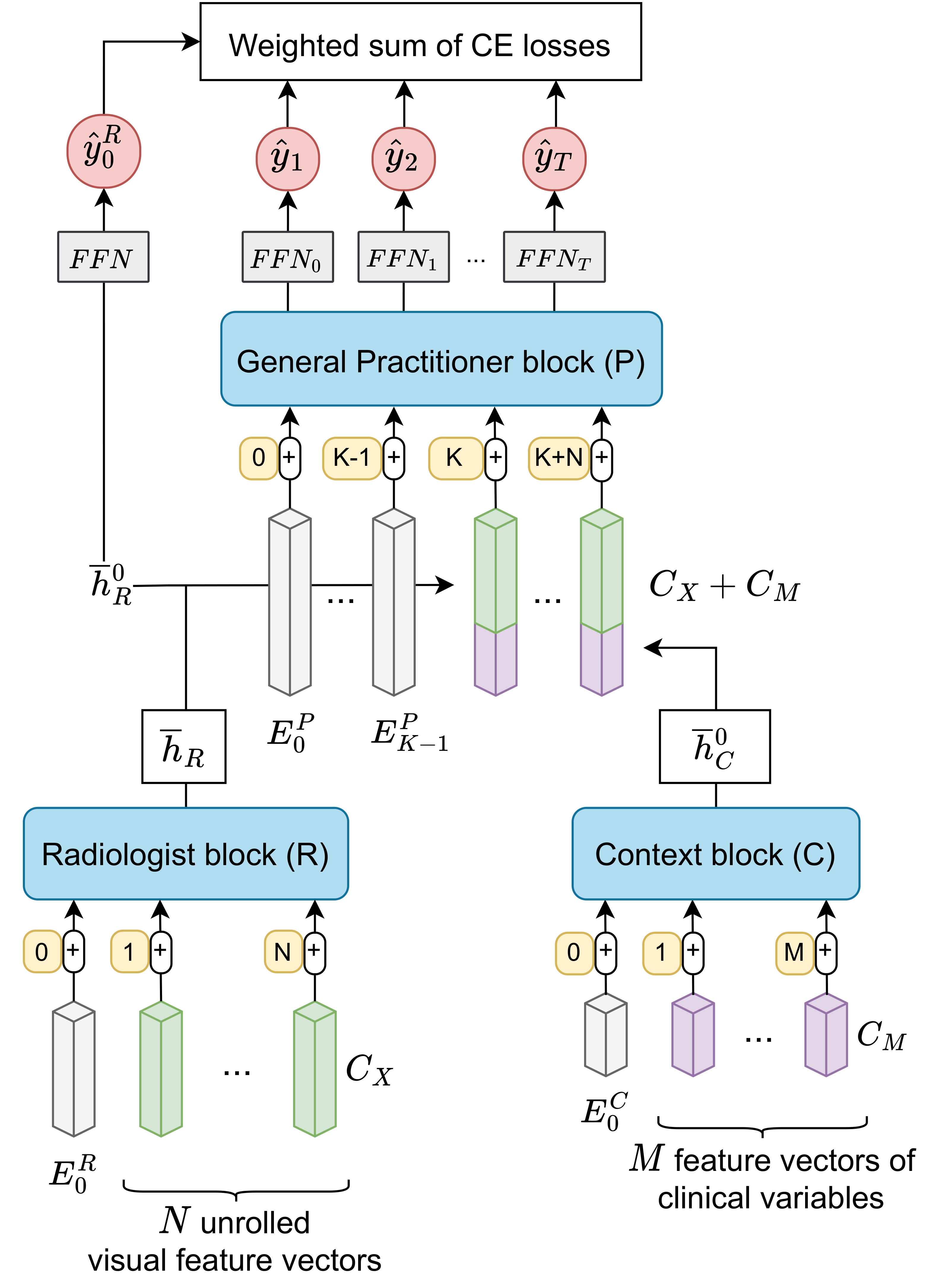}}
    \hfill
    % \subfloat[CLIMATv2]{\includegraphics[height=0.45\textwidth]{figures/CLIMAT/CLIMAT_workflow_v2-crop.pdf}}
    \subfloat[CLIMATv2]{\includegraphics[height=0.45\textwidth]{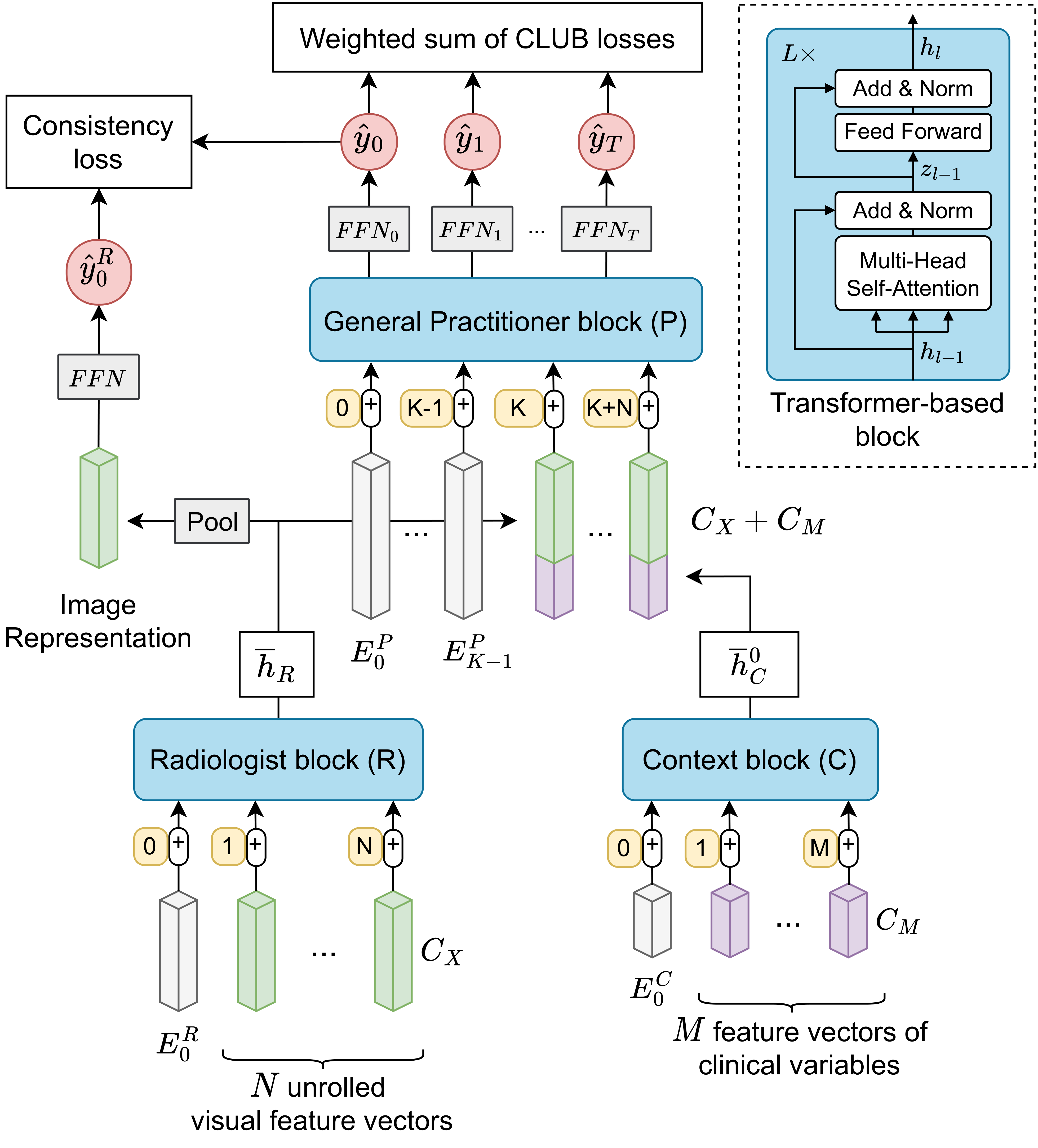}}
    \hspace*{\fill}        
    \caption{\small The CLIMAT framework (best viewed in color). There are $N$ and $M$ input imaging and non-imaging feature vectors, respectively. The first feature vector $\bar{\bm{h}}_C^0$ of the last layer of the transformer C is appended to every output vector of $\bar{\bm{h}}_R$ to form the input for the transformer P. All the blue blocks are transformer-based networks. $[CLS]$ and $[POS]$ embeddings are in white and orange, respectively.}
    \label{fig:climat_workflow}
\end{figure*}

\subsubsection{Transformers for vision tasks}
Although originally developed in the field of natural language processing~\cite{vaswani2017attention,devlin2018bert}, transformer-based architectures have recently been applied also in vision tasks. Dosovitskiy~\etal~\cite{dosovitskiy2020image} pioneered the use of transformer-based architectures without a CNN for image classification problems. Girdhar~\etal~\cite{girdhar2019video} and Arnab~\etal~\cite{arnab2021vivit} studied the same family of architectures to perform video recognition tasks. However, Hassani~\etal~\cite{hassani2021escaping} pointed out such pure transformers require a significantly large amount of imaging data to perform well. The reason is that transformers do not have well-informed inductive biases, which are strengths of CNNs. Thus, our method relies on~\cite{hassani2021escaping} due to medium dataset sizes.

\subsubsection{Multimodal data processing with transformers.}
Transformers have been empirically robust in learning various categories of tasks from sequential data such as text or tabular data~\cite{vaswani2017attention,hu2021transformer}. However, in medical imaging, it is common to acquire multiple modalities comprising both raw images (e.g.\ plain radiographs, MRI, or PET scans) and tabular data, which are challenging for a single transformer. Recent work has shown that multiple transformers are needed to for such multiple modalities~\cite{radford2021learning}. Therefore, similar to our previous version~\cite{nguyen2022climat}, this study adapts the idea of using multiple transformers in our framework to perform DTF from multiple modalities. 

\section{Methods}
\subsection{The CLIMAT framework: a conceptual overview}
\label{sc:conceptual_model}
As mentioned earlier, we base our framework on multi-agent decision-making processes in a clinical setting. In many applications, this can be considered information passing between two agents -- a radiologist and a general practitioner~\cite{jans2013optimizing}. While the radiologist specializing in imaging diagnosis is in charge of producing radiology reports, the general practitioner relies on various modalities including the radiologic findings to forecast the severity of a certain disease. We model such collaboration by the concept presented in~\cref{fig:climat_concept}. Specifically, the radiologist analyzes a medical image $\bm{x}_0$ (e.g.\ radiograph or PET image) of a patient to provide an interpretation with rich visual description and annotations, allowing the diagnosis of the current stage $y_0^R$ of the disease. Subsequently, the general practitioner relies on (i) the clinical data $\bm{m}_0$ (e.g.\ questionnaires or symptomatic assessments) with a further interpretation if needed, (ii) the provided radiology report, and (iii) the referenced diagnosis of the radiologist $y_0^R$ to predict the course of the disease $y_{0:T}$. 

We implement the concept proposed above in the \mbox{CLIMATv2} framework (see~\cref{fig:climat_workflow} and~\cref{sc:climat_framework}). CLIMATv2 comprises three primary transformer-based blocks\footnote{Hereinafter, we use the terms \emph{block} and \emph{transformer} interchangeably.} -- namely Radiologist (R), Context (C), and General Practitioner (P). Firstly, assume that we obtain visual features learned from the imaging data $\bm{x}_0$. Then, the block R acts as the radiologist to perform visual reasoning from the visual features and predict the current stage $\hat y_0^R$ of a disease. The other two blocks are responsible for context extracting and prognosis predicting. As such, the block C aims to extract a context embedding from clinical variables $\bm{m}_0$. Subsequently, the block P utilizes the combination of the context embedding and the output states of the block R to forecast the disease trajectory $\hat y_{0:T}$. 

In this work, we have two major upgrades to CLIMATv1~\cite{nguyen2022climat}. Firstly, we do not assume \updatetwo{anymore}  that $y_0$ and $\bm{m}_0$ are independent, as this does not \removetwo{generally} hold in many medical imaging domains\updatetwo{, e.g. for OA~\cite{liu2023joint}}. \linelabel{ln:version_cmp2}\removetwo{Instead}\updatetwo{Namely, in the current version of CLIMAT}, both the blocks R and P \removetwo{are}\updatetwo{have now been} allowed to make diagnosis predictions simultaneously, 
\updatetwo{making sure that the learned embeddings contain information on $y_0$}. \removetwo{Then}\updatetwo{Furthermore,} we encourage their predictions to be consistent \updatetwo{with the final module of our model}. Secondly, besides performance, \updatetwo{in this work}, we take into account \removetwo{the}\updatetwo{model} calibration\removetwo{aspect}, \updatetwo{which allows us to gain better insights into}\removetwo{corresponding to} the reliability of models' predictions~\cite{guo2017calibration}. \removetwo{For that purpose,}\updatetwo{To facilitate better calibration within our proposed framework,} we propose a novel loss, called CLUB, presented in~\cref{sc:club}.

%\updatetwo{Specifically, Liu~\etal~\cite{liu2023joint} provided empirical evidence of the benefit of the inclusion of non-imaging data in the  grading task. The study conducted by Bird~\etal~\cite{bird2005genetic} indicated a link between human genes and AD while Li~\etal~\cite{li2022validation} showed that a blood test can detect the existence of amyloid-beta plaques in the human brain, which is strongly associated with AD status. In addition, the necessity of non-imaging data was found in the diagnosis of other diseases. In independent studies, Udler~\etal~\cite{udler2019genetic} and Cole~\etal~\cite{cole2020genetics}  showed that genetic data are predictive for diabetes diagnosis. Real-time reverse transcription–polymerase chain reaction (RT-PCR) tests have been practically proven to be the most effective tool for the detection of Coronavirus disease 2019 (COVID-19)~\cite{sule2020real,zhang2021insight}.} 

\subsection{Technical realization}
\label{sc:climat_framework}
\subsubsection{Transformer}
A transformer encoder comprises a stack of $L$ multi-head self-attention layers, whose input is a sequence of vectors $\{\bm{s}_i\}_{i=1}^{N}$ where $\bm{s}_i \in \mathbb{R}^{1\times C}$, and $C$ is the feature size. As such, a transformer is formulated as~\cite{vaswani2017attention}
\begin{align}
    \bm{h}_0 & = [\bm{E}_{[CLS]}, \bm{s}_1, \dots, \bm{s}_N] + \bm{E}_{[POS]}, \label{eq:h_0}\\
    \bm{z}_{l-1} & = \textsc{MSA}(\textsc{LN}(\bm{h}_{l-1})) + \bm{h}_{l-1}, \\
    \bm{h}_{l} & = \textsc{MLP}(\textsc{LN}(\bm{z}_{l-1})) + \bm{z}_{l-1}, \quad l=\{1, \dots, L\} \\
    \bar{\bm{h}} & = \bm{h}_L    
\end{align}
where $\bm{E}_{[CLS]}$ is a learnable token, $\bm{E}_{[POS]}$ is a learnable positional embedding, and $\bar{\bm{h}}$ represents features extracted from the last layer. $\textsc{MLP}$ is a multi-layer perceptron (i.e.\ a fully-connected network), $\textsc{LN}$ is the layer normalization~\cite{ba2016layer}, and $\textsc{MSA}(\cdot)$ is a multi-head self-attention \update{(MSA)} layer~\cite{vaswani2017attention}. 
\linelabel{ln:method_msa}\update{The self-attention mechanism relies on the learning of query, key, value parameter matrices, denoted by $\bm{W}^Q_l, \bm{W}^K_l$, and $\bm{W}^V_l$ with $l=1,\dots,L$, respectively. Initially, we simultaneously set $\bm{Q}_0, \bm{K}_0$, and $\bm{V}_0$ to $\bm{h}_0$ defined in Eq.~\eqref{eq:h_0}. When iterating through layers $l=1,\dots,L$, we update the states as follows
\begin{equation}
\begin{split}
    \bm{Q}_l&=\bm{Q}_{l-1}\bm{W}^Q_l \\ 
    \bm{K}_l&=\bm{K}_{l-1}\bm{W}^K_l \\
    \bm{V}_l&=\bm{V}_{l-1}\bm{W}^V_l
\end{split}
\end{equation}
Finally, the self-attention is established thanks to the \emph{scaled dot-product} function applied to $\bm{Q}_l$, $\bm{K}_l$, and $\bm{V}_l$, and defined as
\begin{equation}
    \mathrm{Attention}(\bm{Q}_l, \bm{K}_l, \bm{V}_l) = \mathrm{Softmax}\left(\frac{\bm{Q}_l\bm{K}_l^\intercal}{\sqrt{d_k}} \right) \bm{V}_l, 
    \label{eq:self_attention_def}
\end{equation}
where $d_k$ is the feature dimension of $\bm{Q}_l$. In essence, $\bm{Q}_l\bm{K}_l^\intercal$ represents the association between all pairs of queries and keys. The normalization based on $d_k$ is critical to address the case where the magnitude of entries in $\bm{Q}_l\bm{K}_l^\intercal$ is too large. The essential part that produces the attention is the utilization of softmax, which allows for the creation of a normalized heatmap over the association of $\bm{Q}_l$ and $\bm{K}_l$. Subsequently, by adding more sets of learnable weights $\bm{W}_l^Q$, $\bm{W}_l^K$, and $\bm{W}_l^V$, we can obtain MSA by concatenating different output heads of attention. Precisely, the MSA mechanism is formulated as follows
\begin{equation}
\begin{split}
    \mathrm{head}^h_l &= \mathrm{Attention}(\bm{Q}_{l}, \bm{K}_{l}, \bm{V}_{l}), \quad h=1,\dots,H
    \\
\mathrm{MSA}(\cdot) &= \mathrm{Concat}(\mathrm{head}^1_l,\dots,\mathrm{head}^H_l)\bm{W}^O_l, \nonumber
\end{split}
\end{equation}
where $H$ is the number of heads, and $\bm{W}_l^O$ represents learning parameters associated with the $H$ output heads.}
% \todo[inline]{How is $d_k$ computed exactly? As described above, it is the feature dim of $Q_l$, and $Q_0 = h_0$.}
% \todo[inline]{Remind the reader what it is in English.}

The three main blocks in our framework are transformer-based networks (see~\cref{fig:climat_workflow}). While the blocks R and C have only $1$ $[CLS]$ token, the block P can include $K~[CLS]$ tokens to allow for multi-target predictions. The hyperparameter $K$ is introduced in the block P to ensure that there are enough output heads for multi-task predictions. We typically set $K$ to $1$ or $T+1$. In the case $K=T+1$, each output head has a corresponding $[CLS]$ token.

\subsubsection{Multimodal feature extraction}
\label{sc:feature_extractor}
Our framework is able to handle multimodal imaging and non-imaging data. As input data can be clinical variables, raw images (i.e.\ 2D or 3D images), and biomarkers extracted by human experts or specialized software, we have distinct feature extraction modules for different input formats. Specifically, we use the feed-forward network (FFN), 2D-CNN, and 3D-CNN-based architectures for scalar or 1D inputs, 2D, and 3D images, respectively. As such, we pre-define common feature lengths $C_X$ and $C_M$ for all imaging and non-imaging embeddings, respectively. Each FFN-based feature extractor consists of a linear layer, a GELU activation~\cite{hendrycks2020gaussian}, layer normalization~\cite{ba2016layer}, and has an output shape of $1\times C_X$ or $1\times C_M$ depending on the type of input data. In the CNN-based modules, we first unroll their output feature maps into sequences of feature vectors per image super-pixel or super-voxel, then linearly project them into a $C_X$-dimensional space. 

\subsubsection{Radiologist module} 
\label{sc:block_r}
The Radiologist block is a transformer network with $L_R$ layers and is responsible for processing all imaging features previously extracted in~\cref{sc:feature_extractor}. For the input data preparation, we concatenate all features of different imaging modalities to form a sequence of length $N$ that contains $C_X$-dimensional image representations. Subsequently, we propagate this sequence through the transformer R.
\linelabel{ln:block_r}\updatetwo{To this end, the visual embedding $\bar{\bm{h}}_R \in \mathbb{R}^{(N+1)\times C_X}$ produced by its last layer serves two purposes: representing radiology reports and visual features for diagnosis predictions. For the former, we subsequently combine $\bar{\bm{h}}_R$ with non-imaging embeddings to constitute inputs for the General Practitioner block (see~\cref{sc:gp_module}). For the latter, following a common practice in~\cite{chu2021conditional,pan2021scalable,park2022vision}, we perform an average pooling onto $\bar{\bm{h}}_R$ to generate a $C_X$-dimensional vector.}
\removetwo{To this end, the visual embedding $\bar{\bm{h}}_R \in \mathbb{R}^{(N+1)\times C_X}$ produced by its last layer is used to model radiology reports. Similar to the clinical setting, $\bar{\bm{h}}_R$ is the means for the Radiologist block to communicate with the General Practitioner block, presented in~}\removecreftwo{\cref{sc:gp_module}}\removetwo{. Furthermore, different from our previous version~\cite{nguyen2022climat}, we allow the block P to make predictions with regard to imaging findings. As such, we first pool $\bar{\bm{h}}_R$ into a $C_X$-dimensional vector by averaging.}
Afterward, we pass the resulting vector through an FFN comprised of a linear layer, a GELU activation~\cite{hendrycks2020gaussian}, and a layer normalization~\cite{ba2016layer} to predict the current stage $y_0^R$ of the disorder (see \cref{fig:climat_workflow}).

\subsubsection{Clinical context embedding module} 
\label{sc:block_c}
Here, we aim to mimic the comprehension of a general practitioner over different clinical modalities (e.g.\ questionnaires, extra tests, and risk factors). As such, we take a single $[CLS]$ embedding followed by $M$ clinical vector representations extracted in~\cref{sc:feature_extractor} to form the input sequence for the Context block (see~\cref{fig:climat_workflow}). The underlying architecture of the block is a transformer-based network. After passing the input sequence through the transformer C with $L_C$ layers, we merely use the first feature vector $\bar{\bm{h}}_C^0$ of the last feature maps $\bm{h}_{L_C}$ as a common contextual token representing all the non-imaging modalities.

\subsubsection{General Practitioner module}
\label{sc:gp_module}
As soon as the contextual token of length $C_M$ is acquired from the Context block, we concatenate \mbox{$N+1$} copies of the token $\bar{\bm{h}}_C^0$ into the last states $\bar{\bm{h}}_R$ of the transformer R to generate a sequence of $N+1$ mixed feature vectors with a feature size of $C_X+C_M$. We then process the obtained sequence using Eq.~\eqref{eq:h_0} to have the sequence of $(K+N+1)$ feature vectors. Here, we utilize the third transformer-based module to simulate the analysis of the general practitioner over all sources of data for prognosis predictions. Specifically, after passing the input sequence through the transformer P, we utilize the first $T+1$ vector representations of the last layer to forecast the disease severity trajectory $(\hat{y}_{0},\dots,\hat{y}_{T})$. Predicting disease severity at each time point requires a common or distinct FFN, which comprises a layer normalization followed by two fully connected layers separated by a GELU activation~\cite{hendrycks2020gaussian}.

\subsection{Calibrated Loss based on Upper Bound for Multi-Task}
\label{sc:club}
\subsubsection{Motivations and formulation}
Compared to CLIMATv1~\cite{nguyen2022climat}, we aim to optimize not only the performance but also the calibration of our model's predictions. As CLIMATv2 simultaneously predicts a sequence of $T+1$ targets with different difficulties, we treat it as a multi-task predictive model. The temporal information here is contained within the transformer states. Inspired by~\cite{kendall2018multi}, to harmonize all the tasks, we propose the CLUB loss (abbreviated from Calibrated Loss Upper Bound). However, unlike~\cite{kendall2018multi}, which relies on the \emph{`not always true'} assumption that $\frac{1}{\sigma} \sum_{c'} \exp \left (\frac{1}{\sigma^2}f_{c'}(x) \right ) = \left ( \sum_{c'} \exp \left ( f_{c'}(x)\right ) \right )^{\frac{1}{\sigma^2}}$, where $\sigma$ is a noise factor and $f_{c'}(.)$ is the $c'$-th element of the logits produced by a parametric function $f$, we theoretically derive CLUB as an upper bound of temperature-scaled cross-entropy (CE) loss.

Consider the $t$-th task with $t \in \{0\dots T\}$, let $\bm{f}_t = (f_{t,1},\dots,f_{t,N_n^t})^\intercal \in \mathbb{R}^{N_c^t}$ denote predicted logits of CLIMATv2 (i.e.\ an output of the transformer P) on task $t$, where $N_c^t$ is the number of classes of the $t$-th target. Let $\bm{g}_{t} = (g_{t,1},\dots,g_{t,N_c^t})^\intercal = \exp(\bm{f}_{t})$. 
Similar to~\cite{kendall2018multi}, we model the affection of noise $\sigma_t$ onto the prediction of $y_t$ in the scaled form $\textsc{Softmax}\left (\frac{1}{\sigma_t^2 + \varepsilon}\bm{f}_t \right )$, where $\varepsilon \in \mathbb{R}_{+}$ is needed to ensure the scaled softmax to be valid for all $\sigma_t \in \mathbb{R}$. For convenience, we temporarily eliminate the $t$ index from all notations. By denoting $\tau=\frac{1}{\sigma^2 + \varepsilon} \in \mathbb{R}_{+}$, we rewrite the scaled softmax as
\begin{equation}
    \textsc{Softmax}\left (\tau\bm{f} \right ) = \left  ( \frac{g_{1}^{\tau}}{\sum_{c'}g_{c'}^{\tau}}, \dots, \frac{g_{N_c}^{\tau}}{\sum_{c'}g_{c'}^{\tau}}\right )^\intercal \in [0, 1]^{N_c},
    \label{eq:softmax}
\end{equation}
where $c$, $c'$ are class indices, and $\tau~\in~\mathbb{R}_{+}$ is a noise factor.

Without the loss of generality, $c$ is assumed to be the ground truth class of a certain input $x$. $\tau$ is the inverse temperature that can smoothen ($\tau \leq 1$) or sharpen ($\tau > 1$) predicted probabilities. Here, one can observe that $\left ( \sum_{c'}g_{c'}^{\tau} \right )^{\frac{1}{\tau}}$ can be seen as an absolutely homogeneous function or an $\ell_\tau$-norm $\left \| \bm{g} \right \|_\tau$ in a Lebesgue space, when $\tau$ belongs to $(0, 1)$ or $[1, \infty)$, respectively. Therefore, a TCE loss can be formulated as
\begin{equation}    
    \mathcal{L}_\mathrm{TCE} = - \log \frac{ g_{c} ^{\tau}}{\left \| \bm{g} \right \|_\tau^\tau},
    \label{eq:TCE_loss}
\end{equation}
where $c$ is the true class. When $\tau=1$, the TCE loss becomes the vanilla CE loss
\begin{equation}    
    \mathcal{L}_\mathrm{CE} = - \log \frac{ g_{c}}{\left \| \bm{g} \right \|_1}.
    \label{eq:CE_loss}
\end{equation}    

For the purpose of improving calibration, we are interested in the case of $\tau \in (0,1]$~\cite{guo2017calibration}, allowing us to apply the reverse Hölder's inequality to have $\left \| \bm{g} \right \|_\tau \leq N_c^{(1-\tau)/\tau}\left \| \bm{g} \right \|_1$. Then, we can derive an upper bound of $\mathcal{L}_\mathrm{TCE}$, called the CLUB loss, as
\begin{equation}
\begin{aligned}
    \mathcal{L}_\mathrm{CLUB} &\triangleq - \tau \log \frac{ g_{c}}{\left \| \bm{g} \right \|_1} + (1 - \tau)\log N_c \nonumber  \\
    &= \tau \mathcal{L}_{\mathrm{CE}} + (1 - \tau) \log N_c,\quad \tau \in [0, 1],    
    \label{eq:CLUB_ub_t<1}
\end{aligned}
\end{equation}
where the equality holds if and only if $\tau = 1$. Unlike $\mathcal{L}_{\mathrm{TCE}}$, our CLUB loss directly depends on $\left \| \bm{g} \right \|_1$ rather than $\left \| \bm{g} \right \|_\tau$. Eq.~\eqref{eq:CLUB_ub_t<1} indicates that $\mathcal{L}_\mathrm{CLUB}$ is a convex combination between the CE loss~\eqref{eq:CE_loss} and $\log N_c$, which takes into account the task complexity in terms of the number of classes.  

\subsubsection{Performance and calibration optimization}
In our setting, we consider each $\tau_t$ associated with task $t$ as a learnable parameter. As the model's parameters $\theta$ and $\tau_t$'s are independent, we can respectively derive the gradients of $\mathcal{L}_\mathrm{CLUB}(t)$ w.r.t. $\theta$ and $\tau_t$'s as follows
\begin{align}
    \frac{\partial \mathcal{L}_\mathrm{CLUB}(t)}{\partial \theta} &= \tau_t\frac{\partial \mathcal{L}_{\mathrm{CE}}(t)}{\partial \theta},\ t=0\dots T, \label{eq:gradient_club_theta} \\    
    \frac{\partial \mathcal{L}_\mathrm{CLUB}(t)}{\partial \tau_t} &= \mathcal{L}_{\mathrm{CE}}(t) - \log N_c,\ t=0\dots T, \label{eq:gradient_club_tau}    
\end{align}
where $\mathcal{L}_{\mathrm{CE}}(t)$ and $\mathcal{L}_\mathrm{CLUB}(t)$ are the CE and CLUB losses on the $t$-th task, respectively. Whereas the optimization w.r.t. $\theta$ essentially aims to improve the performance of our model, learning $\tau_t$'s directly impacts its calibration quality. Eqs.~\eqref{eq:gradient_club_theta} and~\eqref{eq:gradient_club_tau} indicate that $\tau_t$'s can be seen as learnable coefficients of different tasks. 

To effectively constrain $\tau_t \leq 1$ and avoid a trivial solution where $\forall t \in \{0,\dots, T\}, \tau_t = 1$, we constrain the learnable parameters $\{\tau_t\}_{t=0}^T$ using~\cref{alg:CLUB_tau}. Specifically, Line $1$ guarantees that $\rho_t$'s are valid for any $\sigma_t$'s. Lines $2$ to $4$ prevent all the $\tau_t$'s from converging to the obvious value $1$. Lines $5$ and $6$ re-scales $\tau_t$'s such that merely ones with the maximum values become $1$. This last step is necessary to avoid $\tau_t$'s values being small inversely proportionally to the number of tasks.

\subsection{Multi-Task Learning for Disease Trajectory Forecasting}
In practice, it is highly common to have data \emph{not} fully annotated. Thus, our framework should allow for handling missing targets by design. As such, our multi-task loss can tackle such an impaired condition with ease by using an indicator function to mask out targets without annotation. Formally, we minimize the following prognosis forecasting loss
\begin{align}
    \mathcal{L}_{\mathrm{prog}} = \frac{1}{\sum_{t=0}^{T} \mathbb{I}_t} \sum_{t=0}^{T} \mathbb{I}_t  \mathcal{L}_\mathrm{CLUB}(t),      
    \label{eq:l_pn}
\end{align}
where $\mathbb{I}_t$ is an indicator function for task $t$.

\begin{algorithm}[t]
\renewcommand{\arraystretch}{1.75}
\SetAlgoNoLine
\KwInput{$T$: the number of future time points} 
\KwInput{$\{\sigma_t\}_{t=0}^T$: noise parameters}
\KwInput{$\varepsilon\in \mathbb{R}_{+}$: hyperparameter}
    $\rho_t \xleftarrow{} \frac{1}{\sigma^2_t + \varepsilon}$ \quad $t=0\dots T$ \\
    \For{$t = 0,\dots,T$}    
    { 
        $\tilde{\rho}_t = \frac{\exp (\rho_t)}{\sum_{t'=0}^{T} \exp (\rho_{t'})}$ \\
    }    
    $\rho_{\max} \xleftarrow{} \textsc{Max}\left (\{\tilde{\rho}_t\}_{t=0}^T\right )$ \\
    $\tau_t \xleftarrow{} \frac{\tilde{\rho}_t}{\rho_{\max}}$ \quad $t=0\dots T$ \\
 \caption{Constraint of $\tau_t \leq 1$, $t=0\dots T$}
 \label{alg:CLUB_tau}
\end{algorithm}

While the radiologist has strong expertise in imaging diagnosis, in relation to prognosis, the general practitioner has more advantages due to the access to multimodal data, such as the patient's background. On the other hand, general practitioners are also able to assess images to some extent. We incorporate the corresponding prior into our learning framework by enforcing consistency in predictions between the two agents:
\begin{equation}
    \mathcal{L}_{\mathrm{cons}} = \left \| \bm{f}_0^R- \bm{f}_0 \right \|_1, \label{eq:cons_term}
\end{equation}
where $\bm{f}_0^R$ and $\bm{f}_0$ indicate logits of the blocks R and P for diagnosis predictions, respectively. It is worth noting that while $\mathcal{L}_{\mathrm{prog}}$ operates solely on annotated targets, $\mathcal{L}_{\mathrm{cons}}$ optimizes all targets.

To optimize the whole framework, we minimize the final loss $\mathcal{L}$ as follows
\begin{equation}
    \mathcal{L} = \mathcal{L}_{\mathrm{prog}} + \lambda \mathcal{L}_{\mathrm{cons}},
    \label{eq:final_loss}
\end{equation}
where $\lambda\in\mathbb{R}_{+}$ is a  consistency regularization coefficient. 

\section{Experiments}

\subsection{Data}
\begin{figure}[htbp]
    \centering
    \subfloat[OAI dataset]{
    \includegraphics[width=0.48\textwidth]{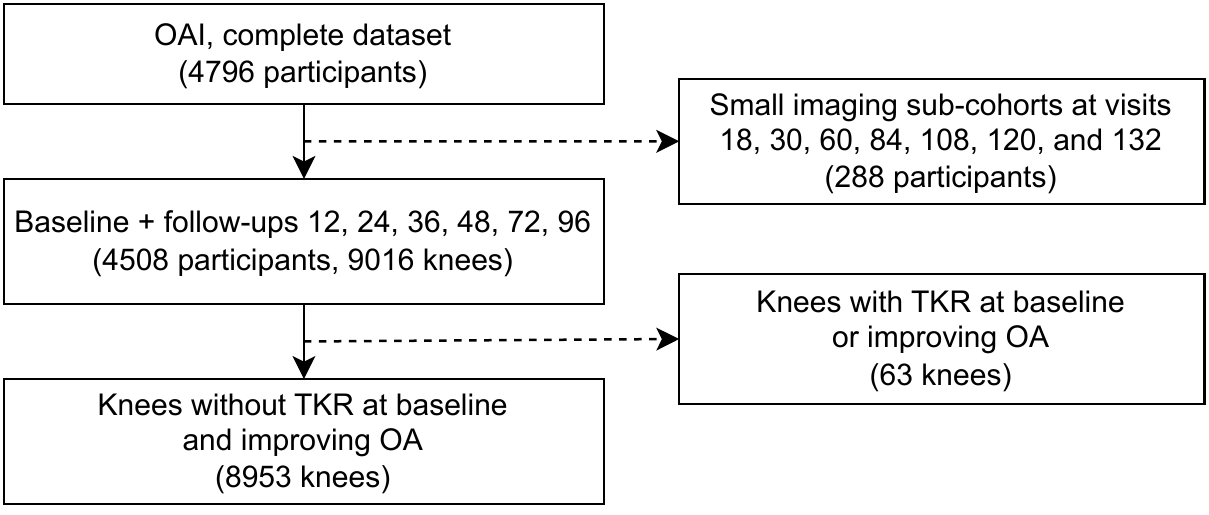}
    }\\
    \subfloat[ADNI dataset]{
    \includegraphics[width=0.48\textwidth]{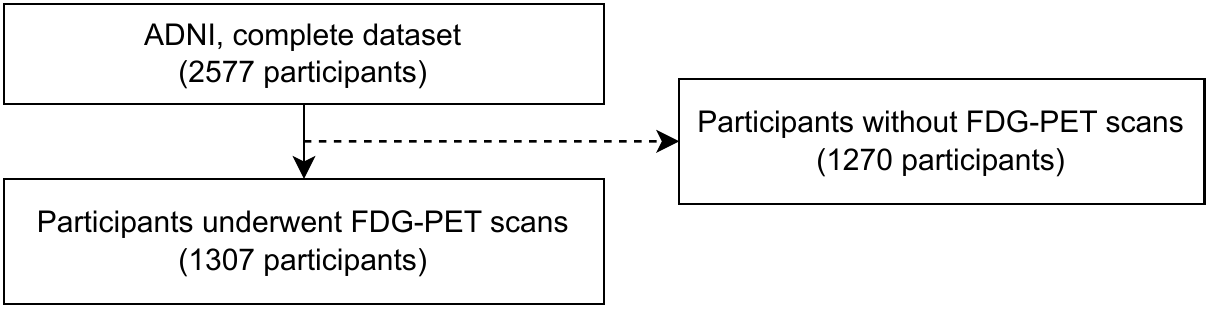}
    }
    \caption{\updatetwo{Subject selection in our study.}}
    \label{fig:subject_selection}
\end{figure}

\linelabel{ln:data_overview}\updatetwo{In this study, we conducted experiments on two public datasets for knee OA and AD. The overall description and subject selection of the two datasets and corresponding tasks can be seen in~\cref{fig:subject_selection,tbl:tasks_desc}. The details of data pre-processing and prognosis prediction tasks are presented as follows.}

\subsubsection{Knee OA structural prognosis prediction}
\label{sc:oai_data}
We conducted experiments on the Osteoarthritis Initiative (OAI) cohort, publicly available at \url{https://nda.nih.gov/oai/}.  $4,796$ participants from $45$ to $79$ years old participated in the OAI cohort, which consisted of a baseline, and follow-up visits after $12$, $18$, $24$, $30$, $36$, $48$, $60$, $72$, $84$, $96$, $108$, $120$, and $132$ months. 
In the present study, we used all knee images that were acquired with large imaging cohorts: the baseline, and the $12$, $24$, $36$, $48$, $72$, and $96$-month follow-ups. 

As the OAI dataset includes data from five acquisition centers, we used data from $4$ centers for training and validation, and considered data from the left-out one as an independent test set. On the former set, we performed a $5$-fold cross-validation strategy.

Following~\cite{tiulpin2019multimodal,tiulpin2020automatic}, we utilized the BoneFinder tool~\cite{lindner2013fully} to extract a pair of knees regions from each bilateral radiograph, and pre-process each of them. Subsequently, we resized each pre-processed image to $256\times256$ pixels \linelabel{ln:oai_pixel_spacing}\update{(pixel spacing of $0.5mm$)}, and horizontally flipped it if that image corresponds to a right knee.

We utilized the Kellgren-Lawrence (KL) as well as OARSI grading systems to assess knee OA severity. The KL system classifies knee OA into $5$ levels from $0$ to $4$, proportional to the OA severity increase. The OARSI system consists of $6$ sub-systems -- namely lateral/medial joint space (JSL/JSM), osteophytes in the lateral/medial side of the femur (OSFL/OSFM), and osteophytes in the lateral/medial side of the tibia (OSTL/OSTM). And according to that the furthest targets in KL, JSL, and JSM were $8$ years from the baseline while it was $4$ years for the other grading aspects.

Regarding the KL grading system, we grouped KL-$0$ and KL-$1$ into the same class as they are clinically similar, and added TKR knees as the fifth class. As a result, there were $5$ classes in KL, and there were $4$ severity levels in each of the OARSI sub-systems. Following~\cite{tiulpin2019multimodal,nguyen2022climat}, we utilized age, sex, body mass index (BMI), history of injury, history of surgery, and total Western Ontario and McMaster Universities Arthritis Index (WOMAC) as clinical variables. We quantized the continuous variables, and presented each of them by a $4$-element one-hot vector depending on the relative position of its value in the interval created by the minimum and the maximum.

For clinical relevance, we did not perform knee OA prognosis predictions on knees that underwent TKR or were diagnosed with the highest grade in any OARSI sub-system. In addition, we ignored one single entry whose pair of knees were improperly localized from its lateral radiograph by the BoneFinder tool. To have more \emph{training} samples, we generated multiple entries from the longitudinal record of each participant by considering imaging and non-imaging data at different follow-up visits (except for the last one) as additional inputs.

\subsubsection{AD clinical status prognosis prediction} 
\label{sc:adni_data}
We applied our framework to forecast the Alzheimer's disease (AD) clinical status from multi-modal data on the Alzheimer's Disease Neuroimaging Initiative (ADNI) cohort, which is available at~\url{https://ida.loni.usc.edu}. The recruitment was done at $57$ sites around America and Canada, and there were $2,577$ male and female participants from $55$ to $90$ enrolled in the cohort. The participants underwent a series of tests such as clinical evaluation, neuropsychological tests, genetic testing, lumbar puncture, MRI, and PET imaging at a baseline and follow-up visits at $1$, $2$, and $4$-year periods.

\begin{table}[t]
\centering
\caption{Dataset statistics. Subjects are \removetwo{the} patient knee joint\updatetwo{s}, and \removetwo{the} patient brain\updatetwo{s} for OAI, and ADNI, respectively.}
\scalebox{1.0}{
\begin{tabular}{llr}
\toprule
\textbf{Dataset} & \textbf{Task} & \multicolumn{1}{l}{\textbf{\# subjects}}  \\ 
\midrule
OAI & OA structural prognosis & 8,953 \\ 
ADNI & AD clinical status prognosis & 1,307 \\ 
\bottomrule
\end{tabular}}
\label{tbl:tasks_desc}
\end{table}

In this study, we used raw FDG-PET scans, MRI measures, cognitive tests, clinical history, and risk factors as predictor variables. The raw FDG-PET scans were pre-processed by the dataset owner,
and were then standardized to voxel dimensions of $160\times160\times160$ \linelabel{ln:exp_vx_std}\update{($1.5\times1.5\times1.5mm^3$ voxel spacing) using the NiBabel library~\cite{brett2020nipy}}. To be in line with the OAI dataset, we applied the same technique to convert scalar inputs to one-hot encoding vectors with a length of $4$. In querying subjects, while we only selected entries whose raw FDG-PET scans were %had to be 
available, the other input variables were allowed to be missing. 

Our objective was to forecast the AD clinical statuses of participants' brains -- cognitively normal (CN), mild cognitive impairment (MCI) or probable AD -- in the next $4$ years. Since the amount of the queried data was substantially limited (see~\linelabel{ln:syn_err1}\updatetwo{\cref{tbl:tasks_desc}}\removecreftwo{ Section IV-A.2}), we sampled entries from follow-up examinations to increase the amount of training data,
and performed $10$-fold cross-validation on this task.

\subsection{Experimental Setup}

\begin{table}[t]
\centering
\caption{\update{Input variables for forecasting knee OA severity grades.}}
\update{
\begin{tabular}{|l|l|l|}
\hline
\textbf{Group} & \textbf{Variable name} & \textbf{Data type} \\ \hline \hline
Raw imaging & Knee X-ray & 2D \\ \hline \hline
Clinical  Variables & Age & Numerical \\ \cline{ 2- 3}
\multicolumn{ 1}{|l|}{} & WOMAC & Numerical \\ \cline{ 2- 3}
\multicolumn{ 1}{|l|}{} & Sex & Categorical \\ \cline{ 2- 3}
\multicolumn{ 1}{|l|}{} & Injury & Categorical \\ \cline{ 2- 3}
\multicolumn{ 1}{|l|}{} & Surgery & Categorical \\ \cline{ 2- 3}
\multicolumn{ 1}{|l|}{} & BMI & Numerical \\ \hline
\end{tabular}}
\label{tbl:oai_input}
\end{table}

\subsubsection{Implementation details}
\label{sc:implementation_details}
We trained and evaluated our method and the reference approaches using V100 \linelabel{ln:exp_gpus}\remove{NVidia}\update{Nvidia} GPUs. \update{Each experimental setting was performed on a single GPU with $12$GB.} We implemented all the methods using the PyTorch framework~\cite{paszke2019pytorch}, 
and trained each of them with the same set of configurations and hyperparameters.
For each problem, we used the Adam optimizer~\cite{kingma2014adam}. The learning rates of $1e\mathrm{-}4$ and $1e\mathrm{-}5$ were set for the OA and AD-related tasks, respectively. 

To extract visual representations of 2D images, we utilized the ResNet$18$ architecture\linelabel{ln:exp_pretrained_resnet}\update{~\cite{he2016deep} whose weights were pretrained on the ImageNet dataset}~\cite{deng2009imagenet}. We used a batch size of $128$ for the knee OA experiments. Regarding 3D images, we chose the 3D-ShuffleNet$2$ architecture because it was well-balanced between efficiency and performance as shown in~\cite{kopuklu2019resource}, which allowed us to train each model with a batch size of $36$ on a single consumer-level GPU. \linelabel{ln:exp_pretrained_3dshufflenet}\update{We utilized 3D-ShuffleNet$2$'s weights previously pretrained on the Kinetics-600 dataset~\cite{carreira2017quo}.\ }Moreover, we used a common feature extraction architecture with a linear layer, a ReLU activation, and a layer normalization~\cite{ba2016layer} for all scalar numerical and categorical inputs. \linelabel{ln:exp_m_n}\update{We provide the detailed description of the input variables in~\cref{tbl:oai_input,tbl:adni_input}.}

\begin{table}[t]
\centering
\caption{\update{Input variables for forecasting clinical statuses of Alzheimer's Disease. See Sec.~\ref{sec:ADClinicalStatusPrognosis} for acronyms.}}
\update{
\begin{tabular}{|l|l|l|}
\hline
\textbf{Group} & \textbf{Variable name} & \textbf{Data type} \\ \hline \hline
Raw imaging & FDG-PET & 3D \\ \hline
\multicolumn{ 1}{|l|}{MRI measures} & Hippocampus & Numerical \\ \cline{ 2- 3}
\multicolumn{ 1}{|l|}{} & Whole brain & Numerical \\ \cline{ 2- 3}
\multicolumn{ 1}{|l|}{} & Entorhinal & Numerical \\ \cline{ 2- 3}
\multicolumn{ 1}{|l|}{} & Fusiform gyrus & Numerical \\ \cline{ 2- 3}
\multicolumn{ 1}{|l|}{} & Mid. temp. gyrus & Numerical \\ \cline{ 2- 3}
\multicolumn{ 1}{|l|}{} & Intracranial vol. & Numerical \\ \hline \hline
\multicolumn{ 1}{|l|}{Clinical variables} & Sex & Categorical \\ \cline{ 2- 3}
\multicolumn{ 1}{|l|}{} & Marriage & Categorical \\ \cline{ 2- 3}
\multicolumn{ 1}{|l|}{} & Race & Categorical \\ \cline{ 2- 3}
\multicolumn{ 1}{|l|}{} & Ethnicity & Categorical \\ \cline{ 2- 3}
\multicolumn{ 1}{|l|}{} & Education & Numerical \\ \hline
\multicolumn{ 1}{|l|}{Cognitive tests} & CDRSB & Numerical \\ \cline{ 2- 3}
\multicolumn{ 1}{|l|}{} & ADAS11 & Numerical \\ \cline{ 2- 3}
\multicolumn{ 1}{|l|}{} & MMSE & Numerical \\ \cline{ 2- 3}
\multicolumn{ 1}{|l|}{} & RAVLT & 1D \\ \hline
\multicolumn{ 1}{|l|}{Cerebrospinal fluid} & A-Beta & Numerical \\ \cline{ 2- 3}
\multicolumn{ 1}{|l|}{} & Tau & Numerical \\ \cline{ 2- 3}
\multicolumn{ 1}{|l|}{} & Ptau & Numerical \\ \cline{ 2- 3}
\multicolumn{ 1}{|l|}{} & Moca & Numerical \\ \cline{ 2- 3}
\multicolumn{ 1}{|l|}{} & Ecog & 1D \\ \hline
\multicolumn{ 1}{|l|}{Risk factors} & Apolipoprotein E4 & Numerical \\ \cline{ 2- 3}
\multicolumn{ 1}{|l|}{} & Age & Numerical \\ \hline
\end{tabular}}
\label{tbl:adni_input}
\end{table}

\subsubsection{Baselines}
For fair comparisons, our baselines were models that had the same feature extraction modules for multi-modal data, as described in~\cref{sc:implementation_details}, but utilized different architectures to perform discrete time series forecasting. As such, we compared our method to baselines with the forecasting module using fully-connected network (FCN), GRU~\cite{cho2014properties}, LSTM~\cite{hochreiter1997long}, multi-modal transformer (MMTF)~\cite{hu2021transformer}, \linelabel{ln:formers_desc}\update{Reformer~\cite{kitaev2020reformer}, Informer~\cite{zhou2021informer}, Autoformer~\cite{chen2021autoformer},} or CLIMAT~\cite{nguyen2022climat}. While FCN, MMTF, \update{Reformer, Informer, Autoformer}, and CLIMAT are parallel models, GRU and LSTM are sequential approaches. \update{Among the transformer-based methods} \remove{Although MMTF, CLIMAT\, and CLIMATv2 are based on the self-attention mechanism,} both versions of CLIMAT have a modular structure of transformers rather than using a flat structure as in MMTF\update{, Reformer, Informer, and Autoformer}. 

\subsubsection{Metrics}
As data from both OAI and ADNI were imbalanced, balanced accuracy (BA)~\cite{brodersen2010balanced} was a must metric in our experiments. As there were only $3$ classes in the AD clinical status prognosis prediction task, we also utilized the one-vs-one multi-class area under the ROC Curve (mAUC-ROC)~\cite{hand2001simple} as another metric. To quantitatively measure calibration, we used expected calibration error (ECE)~\cite{naeini2015obtaining,guo2017calibration}. \linelabel{ln:metric_desc}\update{We reported means and standard errors of each metric computed over $5$ runs with different random seeds.} 

\linelabel{ln:wilcoxon_test_desc}\update{To perform analyses of the statistical significance of our results, we utilized the two-sided Wilcoxon signed-rank test to validate the advantage of our method compared to each baseline~\cite{wilcoxon1992individual}. We equally split the test set into $20$ subsets without overlapping patients. For such a subset, we computed metrics averaged over $5$ random seeds per method. The statistical testing was done patient-wise by comparing our method with every baseline individually. In the case of the OAI dataset, for all patients, we did two rounds of hypothesis testing: one for the left and one for the right knee, respectively. Subsequently, we applied the Bonferroni correction to adjust the significance thresholds for multiple comparisons ($p=0.025$ due to two knees per patient)~\cite{dunn1961multiple}.}

\begin{table}[t]
    \centering
    \renewcommand{\arraystretch}{1.4}
    \caption{Hyperparameter and model selection based on CV performances \updatetwo{on the KL-based knee OA prognosis prediction task}. BA$^{*}$ indicates the averages of BAs of the targets at the baseline and the first $4$ years.}
    \begin{tabular}{cccllc}
\toprule
 Depth & \# of [CLS] &  FFN$_{0:T}$ & $\lambda$  & \updatetwo{Image rep.} & BA$^{*}$ (\%) \\
\midrule \midrule
 2 &   \multirow{3}{*}{9} & \multirow{3}{*}{Separate} &       \multirow{3}{*}{0.5} &    \multirow{3}{*}{\updatetwo{Average pool}} &      59.2 \\
 \textbf{4} &     &  &        &      &     \subbest{59.4} \\
 6 &  &  &        &    &      59.2 \\
   \midrule
\multirow{4}{*}{4} & 1 &   Common &       \multirow{4}{*}{0.5} &  \multirow{4}{*}{\updatetwo{Average pool}}  &         58.3 \\
&\textbf{9} &  \textbf{Common}  &        &        &     \subbest{59.8} \\
& 1 & Separate &      &       &        59.7 \\
& 9 & Separate &        &      &      59.4 \\
\midrule
     \multirow{5}{*}{4} &     \multirow{5}{*}{9} &           \multirow{5}{*}{Common} &       0.0 &   \multirow{5}{*}{\updatetwo{Average pool}}  &   59.1 \\
      &            &    &      0.25 &      &  58.9 \\
      &            &    &       \textbf{0.5} &    &    \subbest{59.8} \\
      &            &    &      0.75 &    &    56.1 \\
      &            &    &       1.0 &    &   56.7 \\
\midrule
\multirow{2}{*}{\updatetwo{4}} & \multirow{2}{*}{\updatetwo{9}} & \multirow{2}{*}{\updatetwo{Common}} & \multirow{2}{*}{\updatetwo{0.5}} & \updatetwo{\subbest{Average pool}} & \updatetwo{\subbest{59.8}} \\
& & & & \updatetwo{[CLS] head} & \updatetwo{58.7} \\
\bottomrule
\end{tabular}
\label{tbl:hyperparam_selection}
\end{table}

\begin{table}[tp]
    \centering
    \caption{\updatetwo{Effect of the consistency term on performance and calibration (K-fold cross-validation). Reported results are averages of BAs and ECEs over the first 4 years.}}
    \updatetwo{\begin{tabular}{lc|cc|cc|cc}
    \toprule
        Grading &  & \multicolumn{2}{c|}{JSL}  & \multicolumn{2}{c|}{JSM} & \multicolumn{2}{c}{AD}  \\ \midrule
             & $\lambda$ & BA & ECE & BA & ECE & BA & ECE \\ \midrule \midrule
        Without $\mathcal{L}_{\mathrm{cons}}$  & 0 & 63.2 & 1.4 & 64.7 & 8.8 & 86.8 & 7.5 \\ \midrule
        \multirow{4}{*}{With $\mathcal{L}_{\mathrm{cons}}$} & 0.25 & 63.1 & 1.5 & 64.5 & 8.7 & 86.8 & 8.3 \\ 
         & 0.50 & 64.9 & 1.8 & 65.3 & 9.6 & 87.3 & 7.9 \\
         & 0.75 & 64.8 & 1.8 & 65.1 & 9.5 & 87.3 & 8.1 \\
         & 1 & 64.6 & 1.7 & 65.2 & 9.5 & 86.7 & 9.2 \\
        \bottomrule
    \end{tabular}}    
    \label{tbl:effect_cons}
\end{table}

\subsection{Ablation studies}
\subsubsection{Overview} We conducted a thorough ablation study to investigate the effects of different components in our CLIMATv2 architecture on the OAI dataset. The empirical results are presented in~\cref{tbl:hyperparam_selection} and summarized in the following subsections.

\subsubsection{Effect of the transformer P's depth} Firstly, we searched for an optimal depth of the transformer P. The results show that the transformer P with a depth of $4$ provides the best performance, yielding $0.2$ \% gain in averaged BA compared to depths of $2$ and $4$. The average BA (over $4$ years) indicates a substantial boost in performance. We, therefore, use the depth of $4$ for the transformer P in the sequel.

\begin{figure}[t]
    \centering
    \croppdf{figures/CLUB/plot_perf_cal_oai_grading_JSM_CLUB_BA}
    \croppdf{figures/CLUB/plot_perf_cal_oai_grading_OSFM_CLUB_BA}
    \croppdf{figures/CLUB/plot_perf_cal_oai_grading_OSTM_CLUB_BA}
    
    \subfloat[Joint space narrowing]{\includegraphics[height=0.15\textwidth]{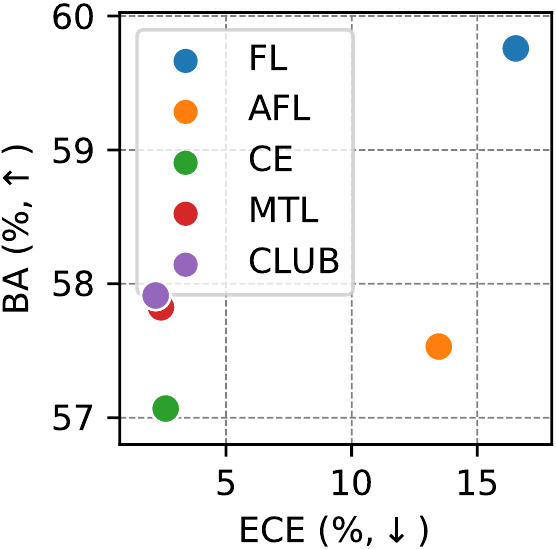}}
    \hfill
    \subfloat[Osteophyte in femur]{\includegraphics[height=0.15\textwidth]{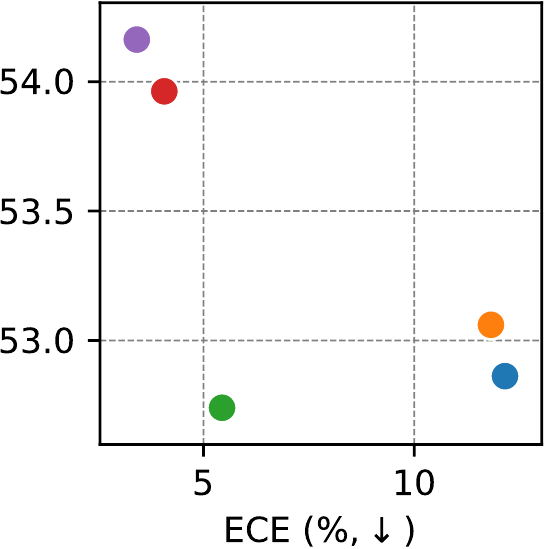}}
    \hfill
    \subfloat[Osteophyte in tibia]{\includegraphics[height=0.15\textwidth]{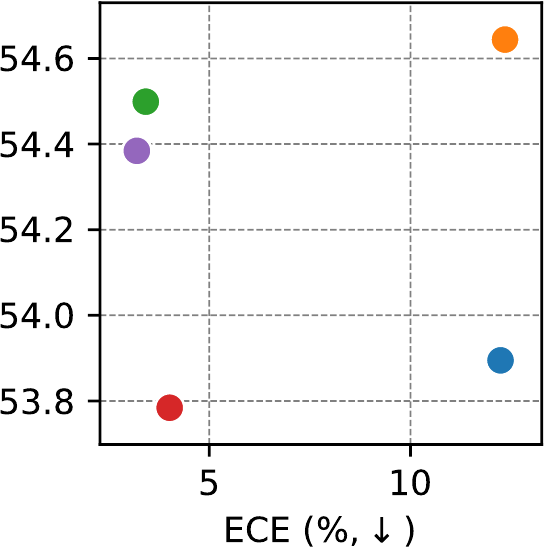}}
    \caption{Performance and calibration comparisons between CLUB and other baselines. All the measures are on the medial side. The losses can be categorized into groups: (1) FL and AFL, and (2) CE, MTL, and CLUB, which are based on cross-entropy and focal loss, respectively.}
    \label{fig:perf_calibration_tradeoff}
\end{figure}

\subsubsection{Effect of the number of [CLS] embeddings and FFNs in the transformer P}
Then, we simultaneously validated two components: using single or multiple [CLS] embeddings, and using common or separate FFN in the transformer P. Of $4$ combinations of settings, the quantitative results suggest that the transformer should have $9$ individual [CLS] embeddings, each of which corresponds to an output head, and merely use one common FFN to make predictions at different time points.

\begin{figure*}[th!]
    % \centering
    \croppdf{figures/DS_OAI/plot_oai_grading_KL_pn_ba}
    \croppdf{figures/DS_OAI/plot_oai_grading_JSM_pn_ba}
    \croppdf{figures/DS_OAI/plot_oai_grading_JSL_pn_ba}
    \croppdf{figures/DS_OAI/plot_oai_grading_OSFM_pn_ba}
    \croppdf{figures/DS_OAI/plot_oai_grading_OSFL_pn_ba}
    \croppdf{figures/DS_OAI/plot_oai_grading_OSTM_pn_ba}
    \croppdf{figures/DS_OAI/plot_oai_grading_OSTL_pn_ba}    
    \subfloat[Kellgren-Lawrence (KL)]{\includegraphics[width=0.33 \textwidth]{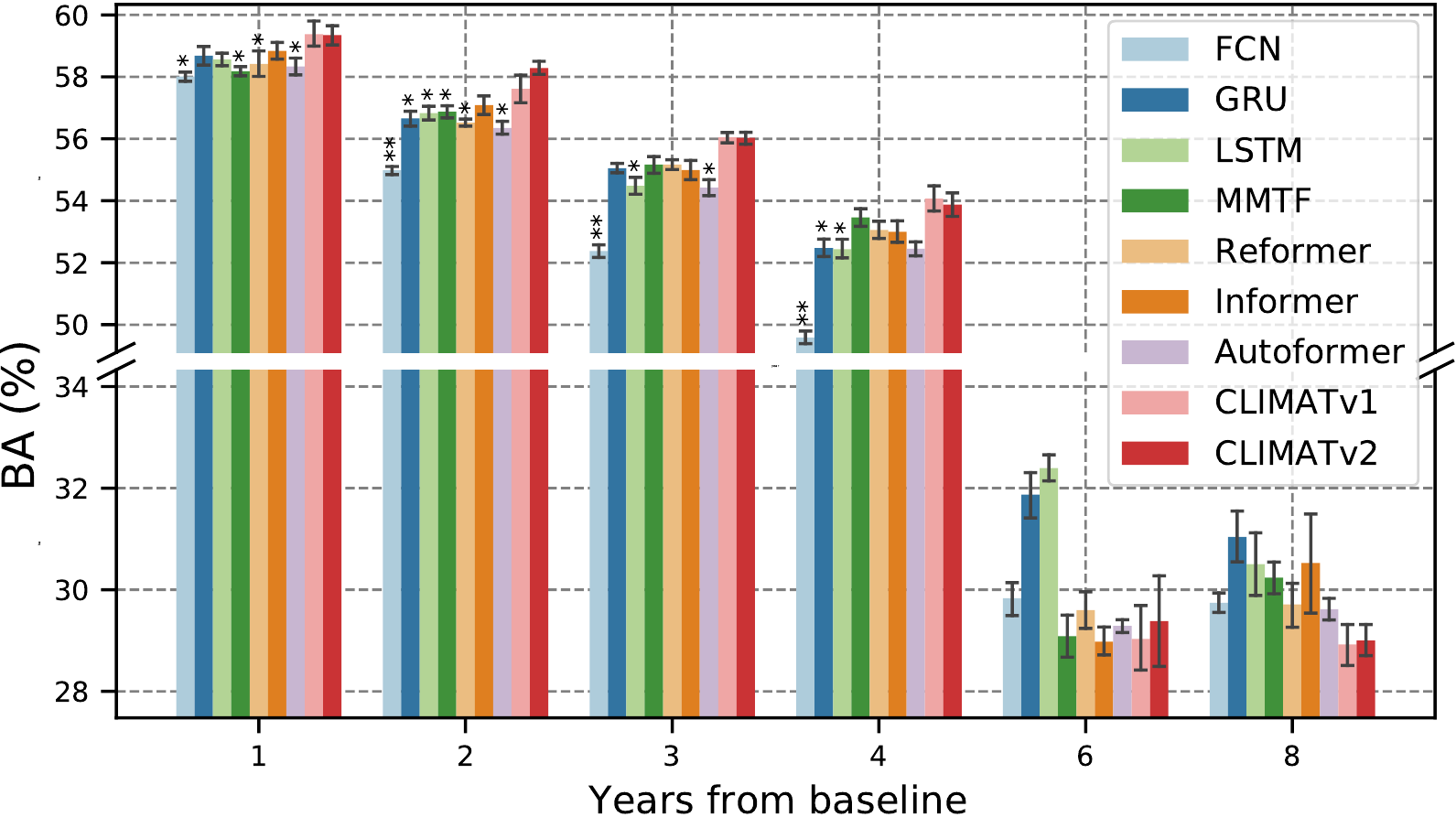}}
    \hfill
    \subfloat[Lateral joint space (JSL)]{\includegraphics[width=0.33 \textwidth]{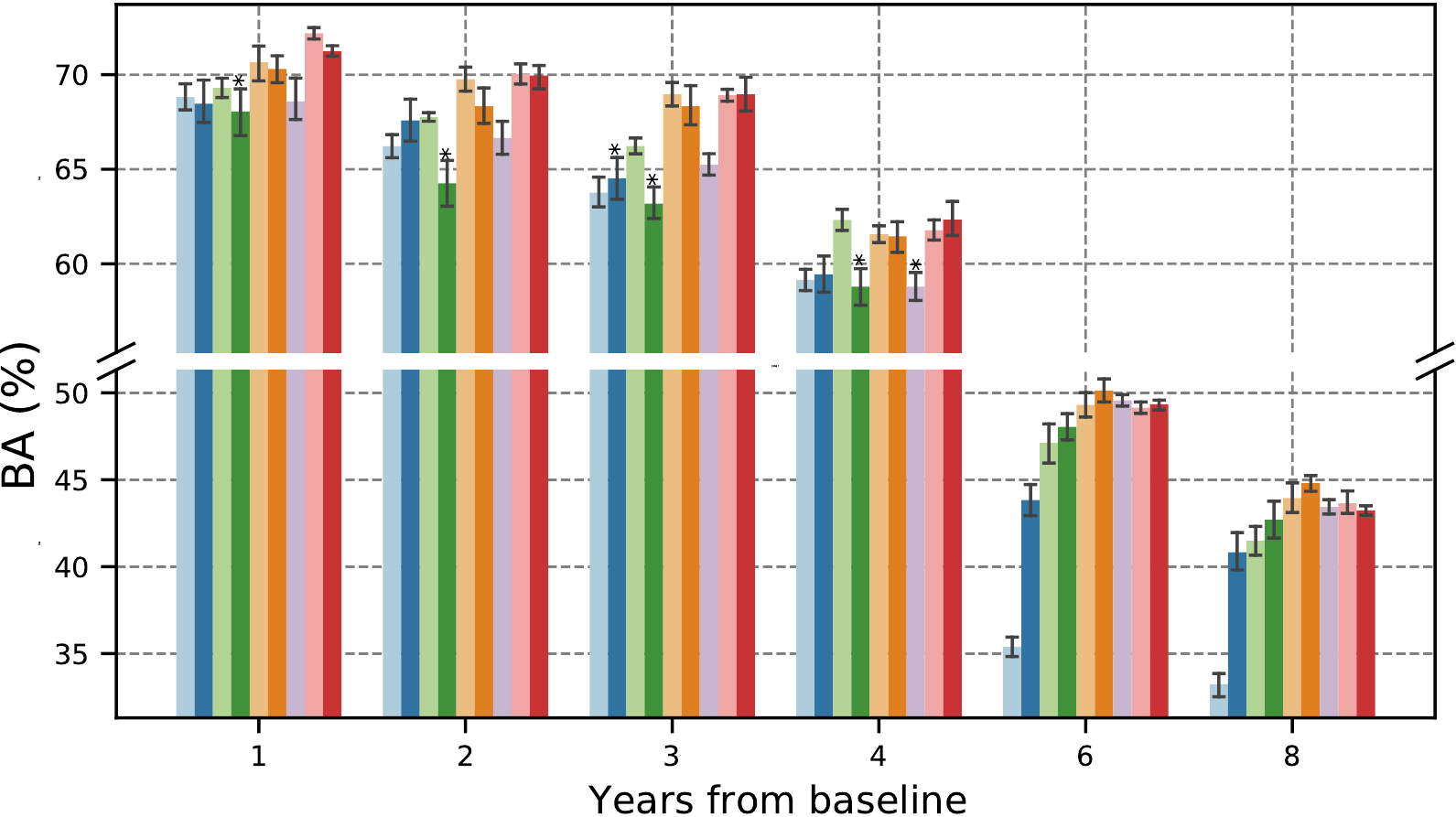}}
    \hfill
    \subfloat[Medial joint space (JSM)]{\includegraphics[width=0.33 \textwidth]{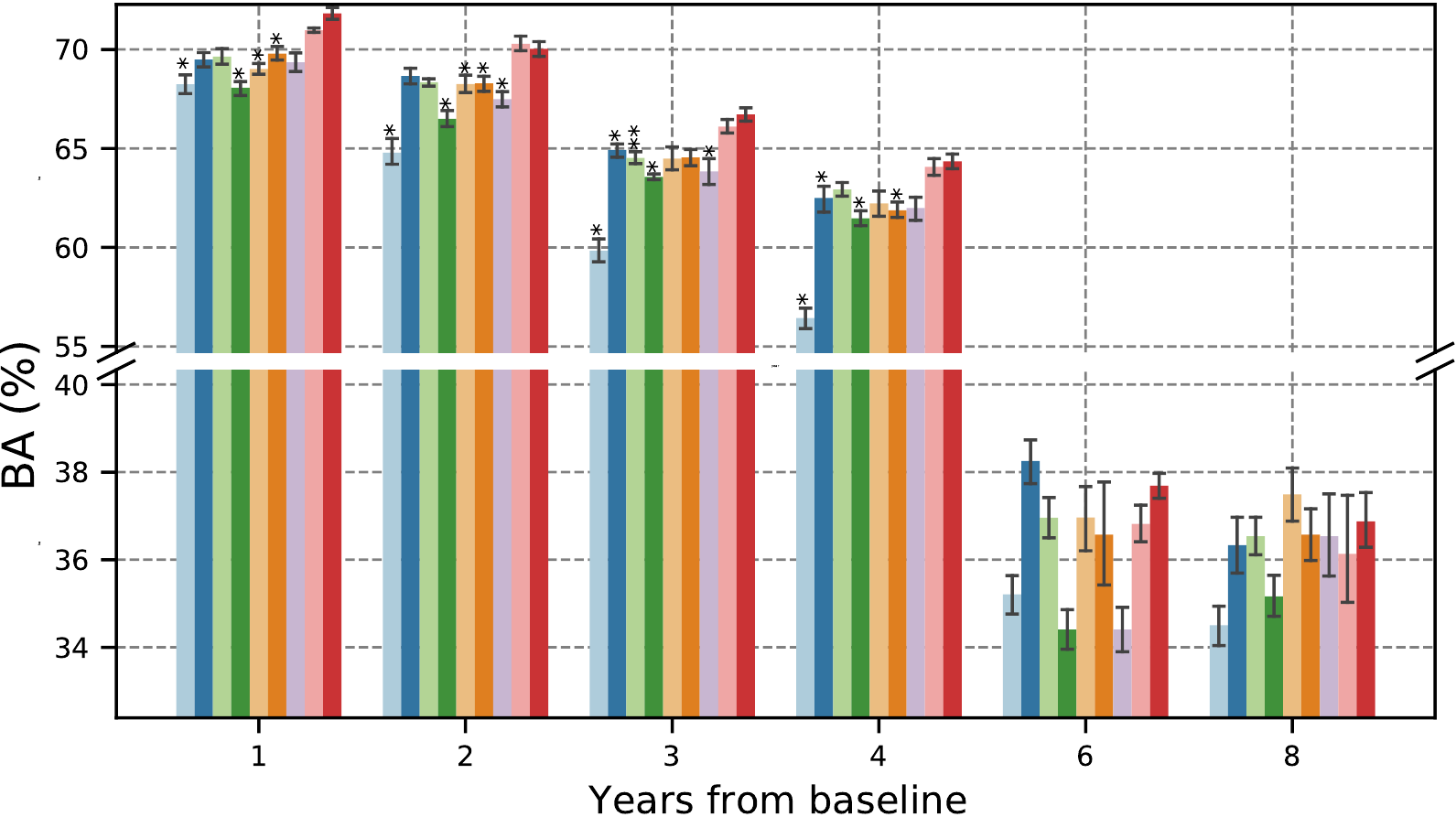}}    
    \\
    \subfloat[Lateral oste. in femur (OSFL)]{\includegraphics[width=0.24 \textwidth]{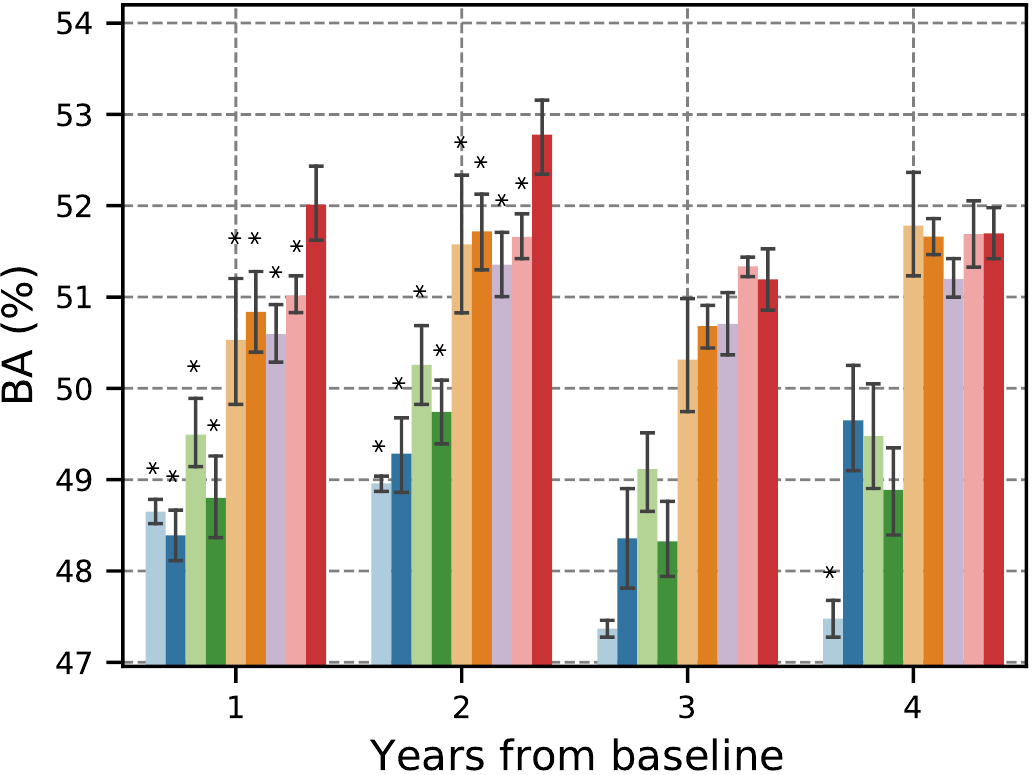}}
    \hfill
    \subfloat[Medial oste. in femur (OSFM)]{\includegraphics[width=0.24 \textwidth]{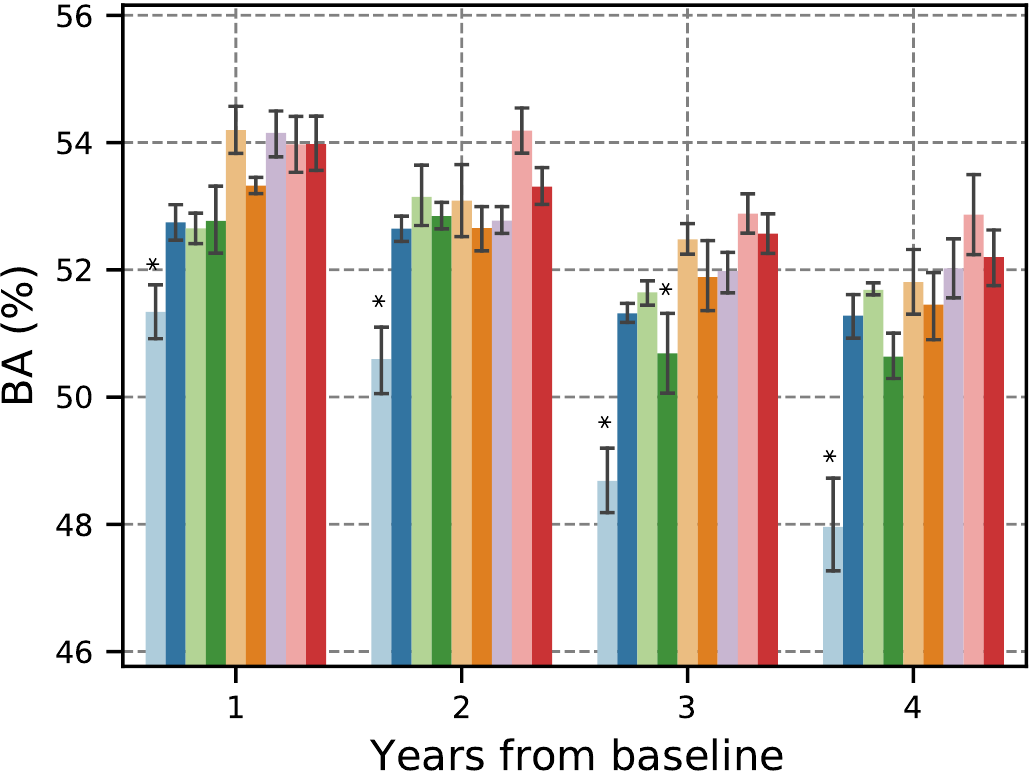}}
    \hfill
    \subfloat[Lateral oste. in tibia (OSTL)]{\includegraphics[width=0.24 \textwidth]{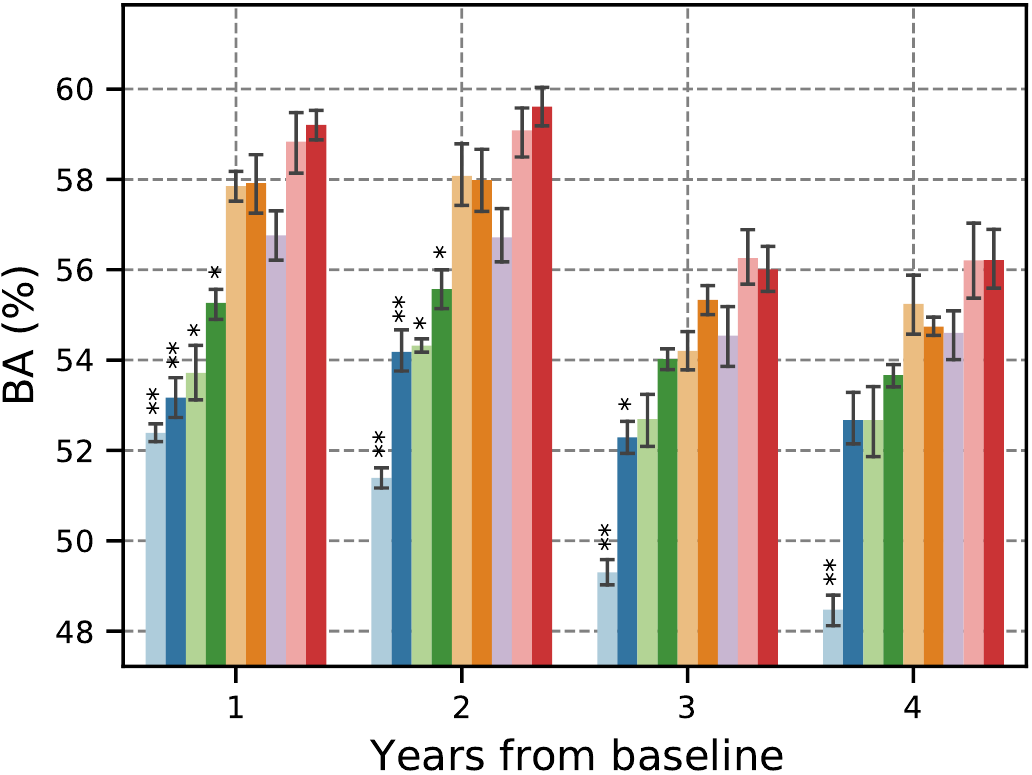}}
    \hfill
    \subfloat[Medial oste. in tibia (OSTM)]{\includegraphics[width=0.24 \textwidth]{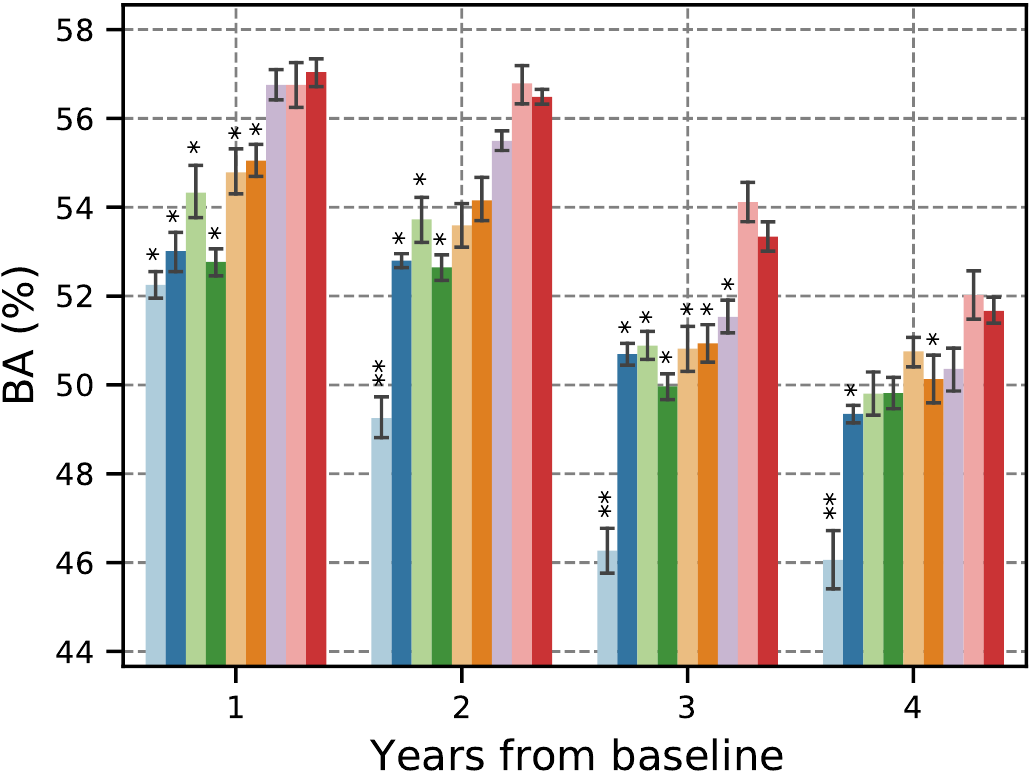}}    
    \caption{\small Performance comparisons between our CLIMAT models and other baselines on the knee osteoarthritis prognosis task via different types of grading (means and standard errors over $5$ random seeds). \update{$*$ and $**$ indicate the statistically significant differences between CLIMATv2 compared to each baseline via Wilcoxon signed-rank tests ($p < 0.05$ and $p < 0.001$, respectively). As the statistical tests were conducted on both knees, p-value thresholds were adjusted to $0.025$ and $0.0005$, respectively.}}
    \label{fig:oai_performances}
\end{figure*}

\subsubsection{Effect of the consistency term}
To validate the necessity of the $\mathcal{L}_{\mathrm{cons}}$ term, we conducted an experiment on a set of $\lambda$ values $\{0, 0.25, 0.5, 0.75, 1\}$. The empirical evidence \linelabel{ln:effect_cons}\updatetwo{in~\cref{tbl:hyperparam_selection}} shows that a $\lambda$ of $0.5$ resulted in the best performance, which was $0.7\%$ higher than the setting without $\mathcal{L}_{\mathrm{cons}}$. \linelabel{ln:effect_cons2}\updatetwo{We further validated the effects of the consistency term on other knee OA grading criteria as well as the AD status forecasting task. The empirical results in~\cref{tbl:effect_cons} consistently demonstrate that the term $\mathcal{L}_{\mathrm{cons}}$ has a positive impact on performance, albeit with the trade-off of calibration. A consistency coefficient $\lambda$ of $0.5$ is the most optimal setting in terms of performance across the tasks. Specifically, we observed BA gains of $1.7\%$, $0.6\%$, and $0.5\%$ with trade-off ECEs of $0.4\%$, $0.8\%$, and $0.4\%$ for JSL, JSM, and AD, respectively.} \removetwo{Thus, the consistency term in~\eqref{eq:final_loss} had a positive effect.}

\subsubsection{Average pooling for image representation}
\label{sc:exp_image_rep}
\linelabel{ln:image_rep}
\updatetwo{In contrast to the previous version, we adopted a conventional approach used in prior studies~\cite{chu2021conditional,pan2021scalable,park2022vision}, which involves performing an average pooling over the output sequence of the Radiologist block to constitute an imaging feature vector for diagnosis prediction $\hat{y}_0^R$. According to~\cref{tbl:hyperparam_selection}, such an approach results in a gain of $1.1\%$ BA compared to the baseline, which solely utilized the first vector of the sequence generated by the block R.}

\begin{figure}[t]
    \centering
    \croppdf{figures/oai_cum_ECE}
    \includegraphics[width=0.49\textwidth]{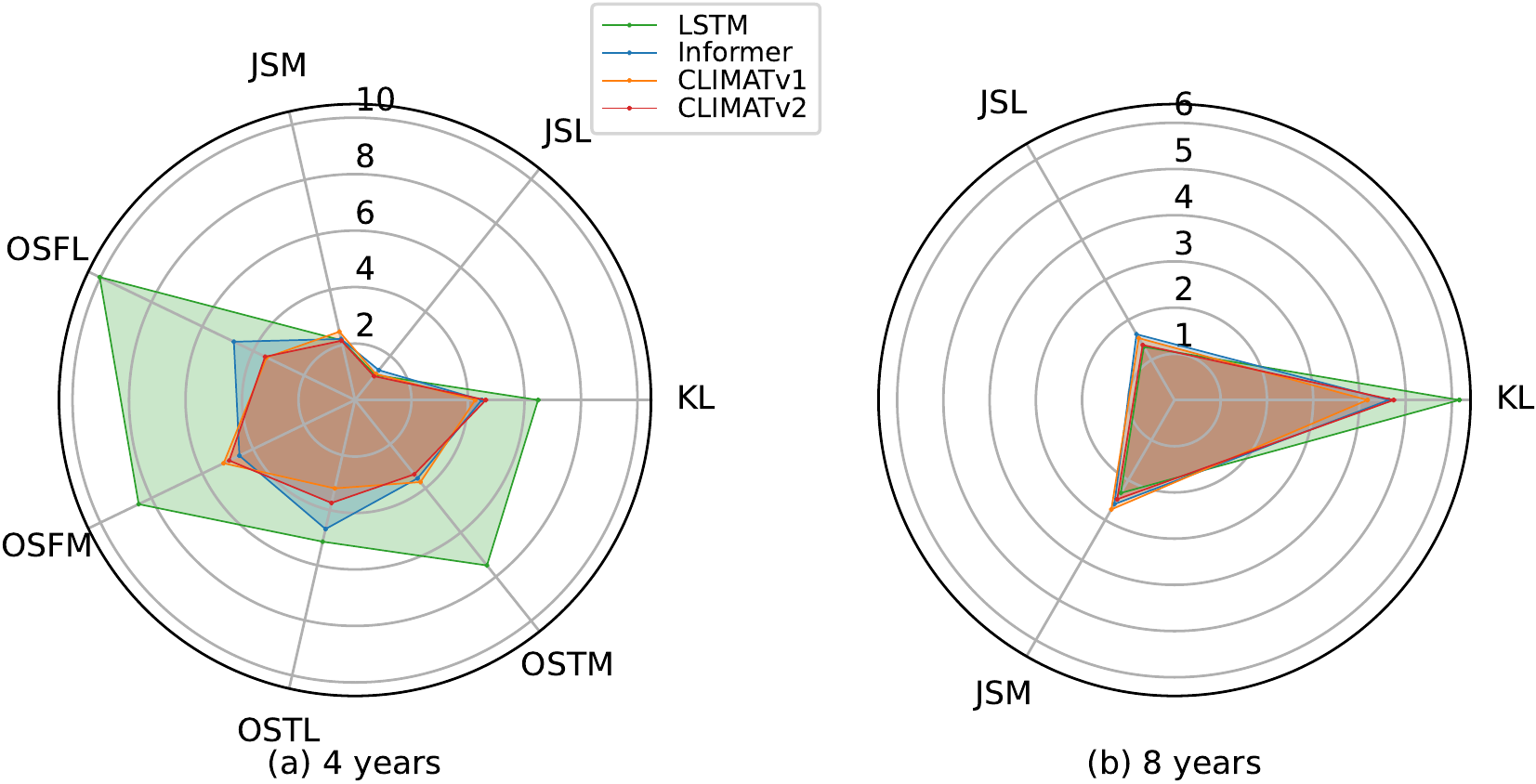}
    \caption{\updatetwo{Calibration comparisons on the knee OA prognosis predictions. (a) Averaged ECEs over the first 4 years. (b) Averaged ECEs over 8 years.}}
    \label{fig:oai_cum_ece}
\end{figure}

\subsubsection{Multimodal channel-wise concatenation}
\label{sc:exp_channelwise_concat}
\updatetwo{We conducted an ablation study on the combination of multimodal embeddings. As such, we compared our channel-wise approach to a sequence-wise baseline that simply concatenates imaging embeddings and a projected version of non-imaging ones. For the baseline, we utilized a linear projection layer to ensure that imaging and non-imaging embeddings are in the same $C_X$-dimensional space. We reported the K-fold cross-validation results in~\cref{tbl:ablation_emb_combination}. On the knee OA-related tasks, our approach tends to have positive benefits on both performance and calibration. Specifically, the performance gains were $2.3\%$, $2.0\%$, and $0.5\%$ for KL, JSL, and JSM, respectively. Except for JSL with an increase of $0.1\%$ ECE, the approach results in calibration improvements of $2.5\%$ and $3.2$ for KL and JSM, respectively. On the AD-related task, the channel-wise approach leads to improvements of $1.1\%$ BA and $0.2\%$ ECE.}

\subsubsection{Effectiveness of CLUB loss}
We compared the CLUB loss to CE itself, multi-task loss (MTL)~\cite{kendall2018multi}, focal loss (FL)~\cite{lin2017focal}, and adaptive focal loss (AFL)~\cite{mukhoti2020calibrating}. Whereas the first two baselines and our loss are based on CE loss, the remaining ones are related to FL. In~\cref{fig:perf_calibration_tradeoff}, we graphically visualize the trade-off between performance and calibration, in which the best in both aspects are expected to locate close to the top-left corners. We observe that our model trained with FL-related losses was substantially worse calibrated compared to the settings with any CE-based loss. Among the losses based on CE, the proposed CLUB helped our model to achieve the best ECEs in all three OA grading systems with insubstantial drops in performance.

\begin{table}[t]
    \centering
    \caption{\updatetwo{Ablation study on imaging and non-imaging combination with k-fold cross-validation ($K=5$ and $K=10$ for OAI and ADNI, respectively). Channel-wise approach (ours) is compared to the sequence-wise approach, concatenating imaging embeddings produced by the block R with projected non-imaging embeddings outputted by the block C. Reported results are averaged BAs and ECEs over the first $4$ years.}}
    \updatetwo{
    \begin{tabular}{llcc}
\toprule
Grading &         Setting &   BA (\%, $\uparrow$) &  ECE (\%, $\downarrow$) \\
\midrule \midrule
    KL &   Channel-wise & \textbf{59.8} & \textbf{14.7} \\
        &  Sequence-wise & 57.5 & 17.2 \\
    \midrule
    JSL &   Channel-wise & \textbf{64.9} &  1.8 \\
        &  Sequence-wise & 62.9 &  \textbf{1.7} \\
    \midrule
    JSM &   Channel-wise & \textbf{65.3} &  \textbf{9.6} \\
        &  Sequence-wise & 64.7 & 12.8 \\
    \midrule
    AD   & Channel-wise & \textbf{87.3} & \textbf{7.9} \\
        & Sequence-wise & 86.2 & 8.1 \\
\bottomrule
\end{tabular}}
    \label{tbl:ablation_emb_combination}
\end{table}

\subsection{Performance and Calibration Comparisons to Competitive Baselines} 
\label{sc:perf_comparisons}
\subsubsection{Knee OA structural prognosis prediction}
In\linelabel{ln:oai_cum_ece}~\cref{fig:oai_performances,fig:oai_cum_ece}, we graphically present comparisons between both versions of CLIMAT and the baselines in the $7$ different knee OA grading scales. 

In general, both versions of CLIMAT outperformed the other baselines in forecasting the knee OA progression within the first $4$ years across the knee OA grading systems. That is consistent with the observation of~\cite{nguyen2022climat} in KL. We observe that LSTM is the most competitive baseline across the grading systems. Compared to LSTM, on average of the first $4$ years, our model achieved $1.3\%$, $1.7\%$, $1.9\%$, $2.3\%$, \linelabel{ln:exp_missing_percent}$0.7\update{\%}$, $4.4\%$, and $2.4\%$ higher BAs while having $1.9\%$, $0.1\%$, $0.02\%$, $6.5\%$, $3.6\%$, $1.4\%$, and $4.1\%$ lower ECEs in KL, JSL, JSM, OSFL, OSFM, OSTL, and OSTM, respectively. \linelabel{ln:exp_oai_former_results}\update{On average, Informer was the most competitive transformer-based baseline. Compared to Informer, CLIMATv2 achieved BA improvements of $0.4\%, 0.3\%, 1.6\%, 0.6\%, 0.4\%, 1.2\%$, and $1.8\%$ in KL, JSL, JSM, OSFL, OSFM, OSTL, and OSTM, respectively. Except for KL and OSFM, our model had lower ECEs with differences of $0.3\%, 0.1\%, 1.5\%, 1.2\%$, and $0.4\%$ in JSL, JSM, OSFL, OSTL, and OSTM, respectively.} Moreover, in comparison to CLIMATv1~\cite{nguyen2022climat}, on average for the first $4$ years, the newer version performed better in KL, JSM, OSFL, and OSTL with BA improvements of $0.1\%$, $0.4\%$, $0.5\%$, and $0.2\%$, respectively, whereas it reached $0.1\%$, $0.5\%$, and $0.3\%$ lower BAs in JSL, OSFM, and OSTM, respectively. Regarding the calibration aspect, CLIMATv2 obtained lower ECEs compared to CLIMATv1 in JSL, JSM, OSFM, and OSTM with differences of $0.1\%$, $0.3\%$, $0.2\%$, and $0.4\%$, respectively.

% \begin{table}[t]
%     \centering
%     \renewcommand{\arraystretch}{1.4}
%     \caption{Hyperparameter and model selection based on CV performances \updatetwo{on the KL-based knee OA prognosis prediction task}. BA$^{*}$ indicates the averages of BAs of the targets at the baseline and the first $4$ years.}
%     \begin{tabular}{ccclc}
% \toprule
%  Depth & \# of [CLS] &  FFN$_{0:T}$ & $\lambda$  &  BA$^{*}$ (\%) \\
% \midrule \midrule
%  2 &   \multirow{3}{*}{9} & \multirow{3}{*}{Separate} &       \multirow{3}{*}{0.5} &          59.2 \\
%  \textbf{4} &     &  &        &           \subbest{59.4} \\
%  6 &  &  &        &           59.2 \\
%    \midrule
% \multirow{4}{*}{4} & 1 &   Common &       \multirow{4}{*}{0.5} &             58.3 \\
% &\textbf{9} &  \textbf{Common}  &        &             \subbest{59.8} \\
% & 1 & Separate &      &               59.7 \\
% & 9 & Separate &        &            59.4 \\
% \midrule
%      \multirow{5}{*}{4} &     \multirow{5}{*}{9} &           \multirow{5}{*}{Common} &       0.0 &        59.1 \\
%       &            &    &      0.25 &        58.9 \\
%       &            &    &       \textbf{0.5} &        \subbest{59.8} \\
%       &            &    &      0.75 &        56.1 \\
%       &            &    &       1.0 &        56.7 \\
% \bottomrule
% \end{tabular}
% \label{tbl:hyperparam_selection}
% \end{table}
% $\hat{y}_0^R$ prediction

\subsubsection{Alzheimer's disease status prognosis prediction}
We reported the quantitative results in~\cref{tbl:adni_performances}. Regarding performance, both the CLIMAT methods achieved the best performances across the prediction targets, in which CLIMATv2 was top-$1$ at the first $2$ years in both BA and mROCAUC. Compared to the transformer-based baseline MMTF, our method outperformed by $2.2\%$, $1.8\%$, and $2.2\%$ BAs at years $1$, $2$, and $4$, respectively. In calibration, CLIMATv2 yielded substantially lower ECEs than all the references at every prediction target. \linelabel{ln:adni_results}\update{That observation was supported by the statistical test results in~\cref{tbl:adni_performances}.}

\begin{table}[t]
    \centering
    \caption{CV performance and calibration comparisons on the ADNI data (mean and standard errors over $5$ random seeds). \update{The best performances with and without substantial differences are indicated by bold and underlined values, respectively. The substantial improvement is determined by whether the best performance overlaps with any other method's. $*$ and $**$ indicate the statistically significant differences between CLIMATv2 vs. each baseline via Wilcoxon signed-rank tests (p < 0.05 and p < 0.001, respectively).}}
    \renewcommand{\arraystretch}{1.4}
\resizebox{0.45\textwidth}{!}{
\begin{tabular}{cllll}
\toprule
\textbf{Year} & \textbf{Method} & \textbf{BA} (\%, $\uparrow$) & \textbf{mROCAUC} (\%, $\uparrow$) & \textbf{ECE} (\%, $\downarrow$) \\
\midrule
\multirow{9}{*}{1} & FCN & 87.6$\pm$0.2$^{**}$ & 96.6$\pm$0.1$^{**}$ & 9.5$\pm$0.3$^{**}$ \\
  & GRU & 87.1$\pm$0.3$^{**}$ & 96.6$\pm$0.1$^{**}$ & 8.6$\pm$0.4$^{*}$ \\
  & LSTM & 87.9$\pm$0.2$^{**}$ & 96.8$\pm$0.1$^{**}$ & 8.7$\pm$0.3$^{**}$ \\
  & MMTF & 88.2$\pm$1.0 & 96.4$\pm$0.8$^{*}$ & 23.3$\pm$0.4$^{**}$ \\
  & Reformer & 80.8$\pm$1.0$^{**}$ & 93.6$\pm$0.7$^{**}$ & 9.6$\pm$0.3$^{**}$ \\
  & Informer & 78.4$\pm$1.5$^{**}$ & 93.1$\pm$0.6$^{**}$ & 10.0$\pm$0.4$^{**}$ \\
  & Autoformer & 85.2$\pm$0.3$^{**}$ & 96.0$\pm$0.1$^{**}$ & 7.8$\pm$0.2$^{*}$ \\  
  \cline{2-5}
  & CLIMATv1 & 90.1$\pm$0.1 & 97.6$\pm$0.1$^{*}$ & 6.7$\pm$0.2 \\
  & CLIMATv2 & \subbest{90.4$\pm$0.1} & \subbest{98.0$\pm$0.1} & \subbest{6.5$\pm$0.1} \\
\midrule
\multirow{9}{*}{2} & FCN & 85.5$\pm$0.2$^{*}$ & 95.7$\pm$0.1$^{*}$ & 9.2$\pm$0.2$^{*}$ \\
  & GRU & 85.1$\pm$0.2$^{*}$ & 95.6$\pm$0.1$^{*}$ & 9.7$\pm$0.2$^{*}$ \\
  & LSTM & 85.7$\pm$0.3$^{*}$ & 95.7$\pm$0.2$^{*}$ & 9.3$\pm$0.7 \\
  & MMTF & 85.4$\pm$0.6 & 95.3$\pm$0.7 & 22.3$\pm$0.4$^{**}$ \\  
  & Reformer & 78.8$\pm$0.9$^{**}$ & 92.5$\pm$0.6$^{**}$ & 9.2$\pm$0.2$^{*}$ \\
  & Informer & 78.1$\pm$1.3$^{**}$ & 92.9$\pm$0.4$^{**}$ & 8.6$\pm$0.1 \\
  & Autoformer & 83.7$\pm$0.3$^{**}$ & 95.2$\pm$0.1$^{**}$ & 8.6$\pm$0.2 \\
  \cline{2-5}
  & CLIMATv1 & 87.2$\pm$0.0 & 96.6$\pm$0.0 & 9.1$\pm$0.2$^{*}$ \\
  & CLIMATv2 & \unsubbest{87.2$\pm$0.2} & \unsubbest{96.6$\pm$0.1} & \subbest{7.9$\pm$0.3} \\
\midrule
\multirow{9}{*}{4} & FCN & 80.7$\pm$0.3$^{*}$ & 93.7$\pm$0.1 & 9.6$\pm$0.3$^{*}$ \\
  & GRU & 81.7$\pm$0.1 & 93.8$\pm$0.1 & 12.6$\pm$0.4$^{**}$ \\
  & LSTM & 81.0$\pm$0.7$^{*}$ & 93.5$\pm$0.2 & 12.4$\pm$0.6$^{**}$ \\
  & MMTF & 80.6$\pm$0.7 & 93.0$\pm$0.6 & 17.9$\pm$0.3$^{**}$ \\
  & Reformer & 76.7$\pm$0.9$^{**}$ & 90.5$\pm$0.5$^{**}$ & 11.1$\pm$0.5$^{*}$ \\
  & Informer & 71.6$\pm$0.9$^{**}$ & 87.4$\pm$0.3$^{**}$ & 12.6$\pm$0.3$^{**}$ \\
  & Autoformer & 80.5$\pm$0.3$^{**}$ & 93.3$\pm$0.1$^{**}$ & 9.5$\pm$0.2 \\
  \cline{2-5}
  & CLIMATv1 & \unsubbest{83.0$\pm$0.2} & \subbest{94.5$\pm$0.1} & 9.6$\pm$0.3$^{*}$ \\
  & CLIMATv2 & 82.8$\pm$0.1 & 94.2$\pm$0.1 & \subbest{9.2$\pm$0.2} \\
\bottomrule
\end{tabular}
\label{tbl:adni_performances}
}
\end{table}

\subsection{Attention maps over multiple modalities}

The self-attention mechanism of the transformers in CLIMATv2 allowed us to visualize attention maps over imaging and non-imaging modalities when our model made a prediction at a specific target. Specifically, we used $\mathrm{Softmax}\left(\bm{Q}_L\bm{K}_L^\intercal / \sqrt{d_k} \right)$, where $\bm{Q}_L, \bm{K}_L$ are query and key matrices of the last layer $L$, respectively, and $d_k$ is the feature dimension of the key matrix, as attention maps~\cite{vaswani2017attention}. While we utilized the softmax output corresponding to $\bar{h}_C^0$ in the transformer F for clinical variables, we took the softmax output in computing $\bar{h}_P^t$ with $t=0,\dots,T$ in the transformer P to visualize attention maps on imaging modalities. Here, we set $t=1$, corresponding to the forecast of a disease severity $1$ year from the baseline.

\subsubsection{Knee OA structural prognosis}
\begin{figure}[t!]
    \centering    
    \croppdf{figures/Attention_map/KL_OARSI/batch_C_02_knee_49_all_0.0}
    \croppdf{figures/Attention_map/KL_OARSI/batch_C_02_meta_49_all_0.0}
    \croppdf{figures/Attention_map/KL_OARSI/batch_C_02_preds_49_all_0.0}    
        \hspace*{\fill}
        \subfloat[A knee with attention maps]{
        \includegraphics[width=0.2\textwidth]{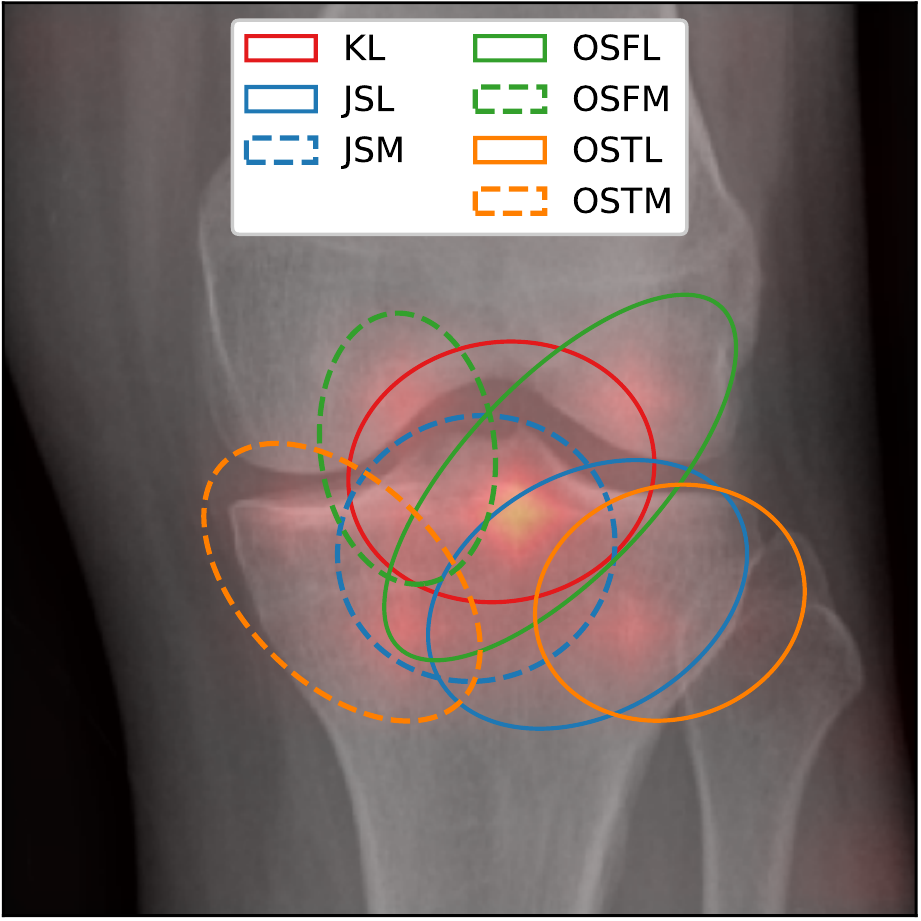}
        \label{fig:oai_img_interpretability}}
        \hfill
        \subfloat[Contributions of the variables]{
        \includegraphics[width=0.21\textwidth]{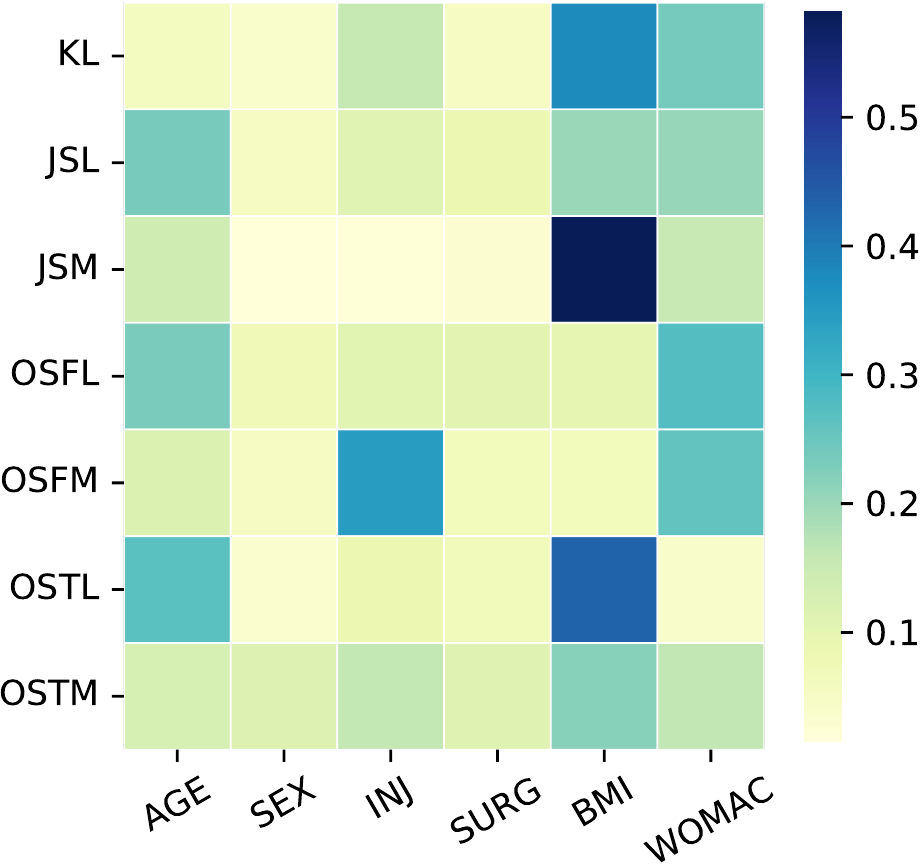}
        \label{fig:oai_meta_interpretability}}  
        \hspace*{\fill}        
    \caption{\small An example of progression from a healthy knee at baseline to early osteoarthritis. Our model identified the changes in the intercondylar notch, sex, and symptomatic status.}
    \label{fig:oai_interpretability}
\end{figure}

In~\cref{fig:oai_interpretability}, we visualized attention maps over different input modalities across $7$ grading criteria. As such, in~\cref{fig:oai_img_interpretability}, we displayed a healthy knee at the baseline overlaid by $7$ corresponding saliency maps. For differentiation, we also provided colored ellipses. \cref{fig:oai_meta_interpretability} shows the heatmap over the $6$ clinical variables on each grading criterion. Values on each row sum up to $1$. In this particular case, we observe that the model has paid the most attention to the intercondylar notch, together with BMI and WOMAC~\cite{leon2005intercondylar}. 

\subsubsection{AD clinical status prognosis}\label{sec:ADClinicalStatusPrognosis}
\begin{figure}[t!]
    \centering
    % \IfFileExists{figures/figures/ADNI_attn_map/MCI/batch_3_1_1_3_imaging-crop.pdf}{}{\immediate\write18{pdfcrop figures/ADNI_attn_map/MCI/batch_3_1_1_3_imaging.pdf}}
    % \IfFileExists{figures/ADNI_attn_map/MCI/batch_3_1_1_3_meta-crop.pdf}{}{\immediate\write18{pdfcrop figures/ADNI_attn_map/MCI/batch_3_1_1_3_meta.pdf}}
    % \croppdf{figures/ADNI_attn_map/MCI/batch_1_06_diag1_np_4_pet_axial}
    % \croppdf{figures/ADNI_attn_map/MCI/batch_1_06_diag1_np_4_imaging}
    % \croppdf{figures/ADNI_attn_map/MCI/batch_1_06_diag1_np_4_meta}    
    \croppdf{figures/ADNI_attn_map/list/batch_1_00_diag1_np_7_pet_axial}
    \croppdf{figures/ADNI_attn_map/list/batch_1_00_diag1_np_7_imaging}
    \croppdf{figures/ADNI_attn_map/list/batch_1_00_diag1_np_7_meta}
    \subfloat[Attention maps on an axial FDG-PET slice]{
        \includegraphics[trim={7.5cm, 0 7.5cm, 0},clip, width=0.18\textwidth]{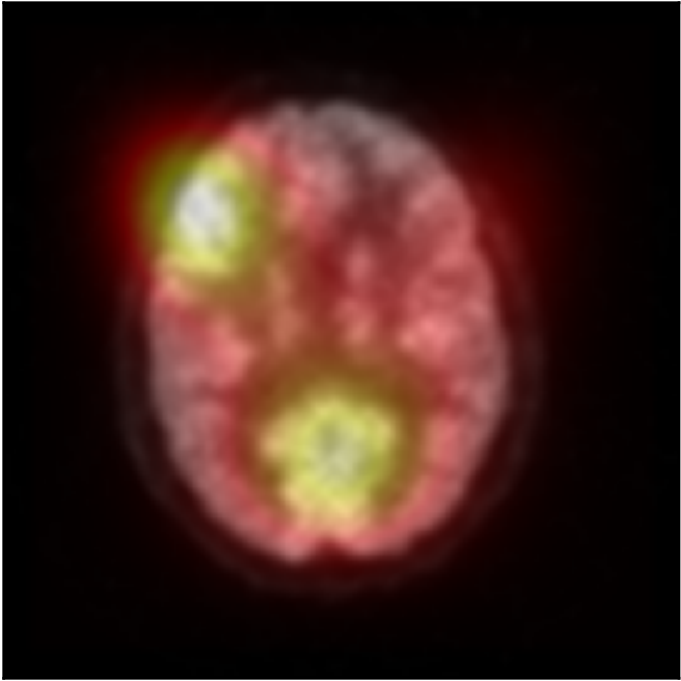}
        \label{fig:adni_img_interpretability}} \hfill
        \subfloat[FDG-PET and imaging variables]{
        \includegraphics[width=0.24\textwidth]{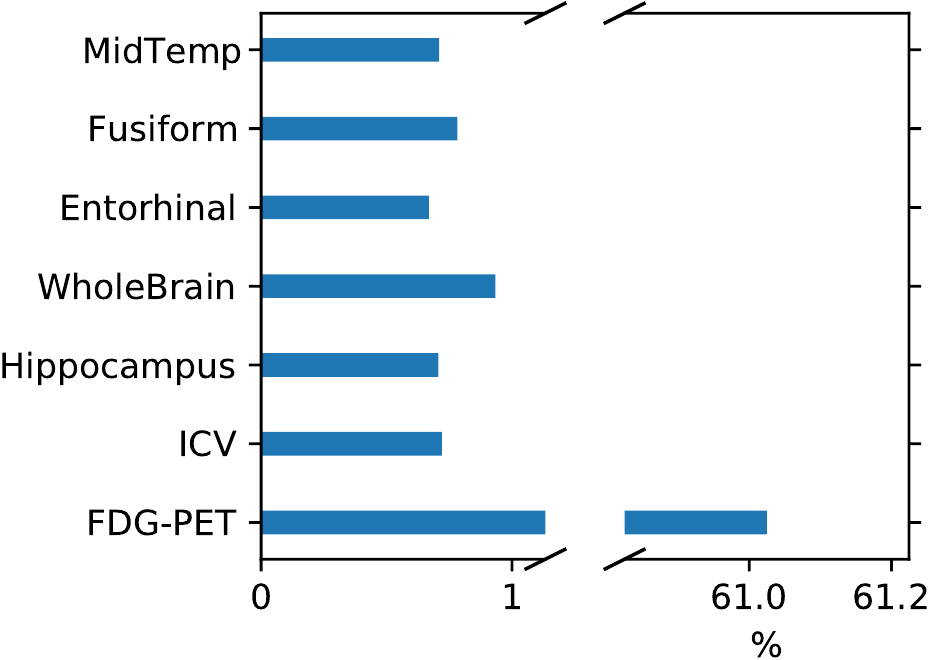}
        \label{fig:adni_all_img_interpretability}} 
        \\        
        \subfloat[Importances of other variables]{
        \includegraphics[width=.47\textwidth]{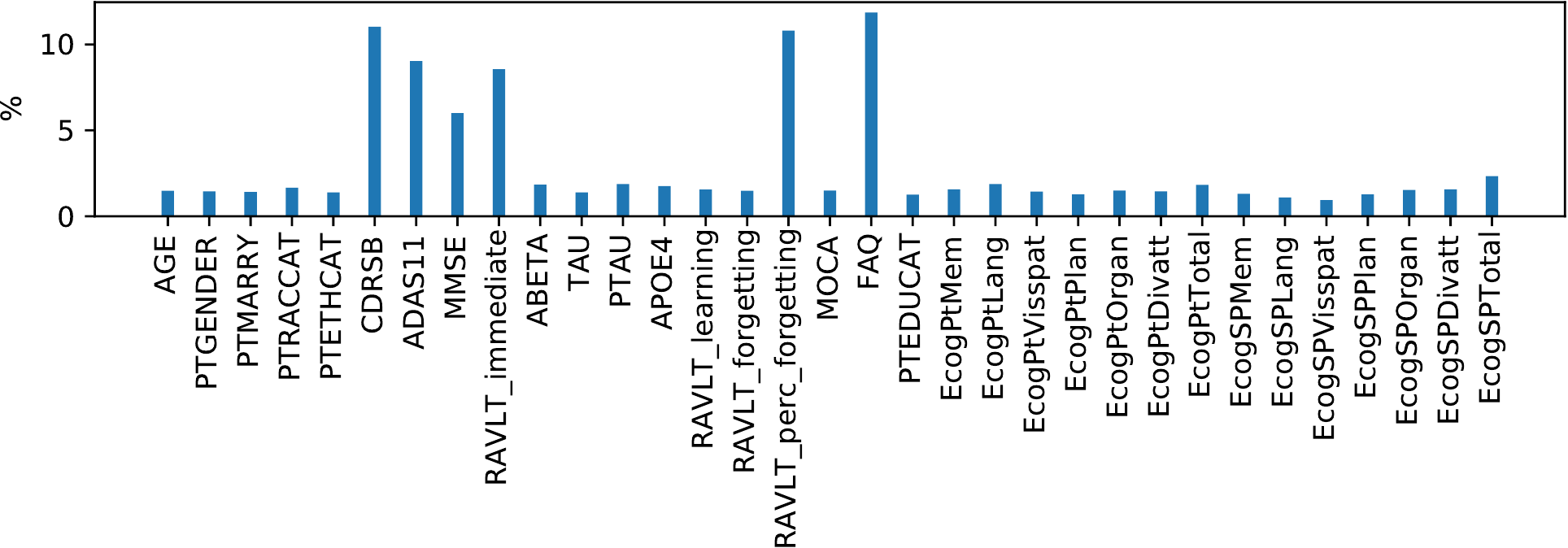}
        \label{fig:adni_meta_interpretability}} 
    \caption{\small Interpretability of our method's prediction on a \update{selective} sample from the ADNI dataset.}
    \label{fig:adni_interpretability}
\end{figure}

As imaging data consisted of 3D FDG-PET scans as well as the other imaging measurements, we had to separate them into~\cref{fig:adni_img_interpretability,fig:adni_all_img_interpretability}. We can observe that an attention sphere locates around \linelabel{ln:adni_attn}\update{the posterior cingulate cortex, }the inferior frontal gyrus, and the middle gyrus~\cite{hallam2020neural}.~\cref{fig:adni_all_img_interpretability} shows accumulated attention weights corresponding to the FDG-PET feature vectors alongside ones of the other imaging measurements. The reason \linelabel{ln:adni_attn_result}\remove{for that}\update{that imaging variables were assigned a substantially higher importance} is that the number of the 3D visual embeddings was dominant compared to the others (i.e.\ $125$ versus $6$). In~\cref{fig:adni_meta_interpretability}, high attention can be observed on the percent forgetting score of the Rey Auditory Verbal Learning Test (RAVLT), RAVLT immediate, the AD assessment score 11-item (ADAS11), Clinical Dementia Rating Scale–Sum of Boxes (CDRSB), Mini-Mental State Exam (MMSE), and Functional Activities Questionnaire (FAQ).
%\updatetwo{Specifically, Liu~\etal~\cite{liu2023joint} provided empirical evidence of the benefit of the inclusion of non-imaging data in the knee OA grading task. The study conducted by Bird~\etal~\cite{bird2005genetic} indicated a link between human genes and AD while Li~\etal~\cite{li2022validation} showed that a blood test can detect the existence of amyloid-beta plaques in the human brain, which is strongly associated with AD status. In addition, the necessity of non-imaging data was found in the diagnosis of other diseases. In independent studies, Udler~\etal~\cite{udler2019genetic} and Cole~\etal~\cite{cole2020genetics}  showed that genetic data are predictive for diabetes diagnosis. Real-time reverse transcription–polymerase chain reaction (RT-PCR) tests have been practically proven to be the most effective tool for the detection of Coronavirus disease 2019 (COVID-19)~\cite{sule2020real,zhang2021insight}.} 

\section{%Discussion and 
Conclusions}
\label{sc:conclusions}
In this paper, we proposed a novel general-purpose transformer-based method to forecast the trajectory of a disease's stage from multimodel data. We applied our method to two real-world applications, that are related to OA and AD. 
Our framework provides tools to integrate multi-modal data and has interpretation capabilities through self-attention. \linelabel{ln:conclusion_optimization}\update{\removetwo{As an upgrade of CLIMATv1}\removecreftwo{~\cite{nguyen2022climat}}\removetwo{, the framework allows us to optimize for both performance and calibration on the disease forecasting task.}}
\linelabel{ln:attn_limitation_removed}\remove{It is noteworthy that attention maps produced by transformers act as human-friendly signals of our model, and should be carefully used in practice with expert knowledge in the domain.} \remove{Transformers may highlight areas not associated with the body part, which can be seen in~}\removecref{\cref{fig:adni_img_interpretability}}\remove{, as well as in prior art~\cite{petit2021u,odusami2022intelligent}.}\note{(moved to Line~\ref{ln:attn_limitation})}\notetwo{(moved to Line~\ref{ln:summary})} \linelabel{ln:summary_prev}\removetwo{To our knowledge, this is not only the first study in the realm of OA, but also the first work on AD clinical status prognosis prediction from the multi-modal setup that includes raw 3D scans and scalar variables. The developed method can be of interest to other fields, where forecasting of disease course is of interest.} 

\linelabel{ln:conclusion_climat_cmp}\updatetwo{In comparison with the prior version, CLIMATv2 has two primary upgrades. First, we have eliminated the assumption of independence between non-imaging data $m_0$ and diagnostic predictions $y_0$ used in CLIMATv1~\cite{nguyen2022climat} since it does not hold not only in OA and AD, but also in other diseases. Specifically, Liu~\etal~\cite{liu2023joint} provided empirical evidence of the benefit of the inclusion of non-imaging data in the knee OA grading task. The study conducted by Bird~\etal~\cite{bird2005genetic} indicated a link between human genes and AD while Li~\etal~\cite{li2022validation} showed that a blood test can detect the existence of amyloid-beta plaques in the human brain, which is strongly associated with AD status. %In addition, the necessity of non-imaging data was found in the diagnosis of other diseases. In independent studies, Udler~\etal~\cite{udler2019genetic} and Cole~\etal~\cite{cole2020genetics} showed that genetic data are predictive for diabetes diagnosis. Real-time reverse transcription–polymerase chain reaction (RT-PCR) tests have been practically proven to be the most effective tool for the detection of Coronavirus disease 2019 (COVID-19)~\cite{sule2020real,zhang2021insight}. 
Second, we have proposed the CLUB loss, which allowed us to optimize for both performance and calibration.}

\linelabel{ln:cons_limitations}\update{There are some limitations in this study, which are worth mentioning. First, we used common DL architectures as imaging and non-imaging feature extractors. While such a standardized procedure resulted in fair comparisons, better results could have been obtained with e.g. Neural Architecture Search methods~\cite{elsken2019neural}. Furthermore, a wider range of DL modules could have been considered, but this could substantially increase the use of computing resources. Specifically, to obtain results in this work, it required roughly 400 GPU hours for experiments in~\cref{tbl:adni_performances} and 525 GPU hours in~\cref{fig:oai_performances} for every method, respectively.} 

\begin{figure}
    \centering
    \includegraphics[width=.47\textwidth]{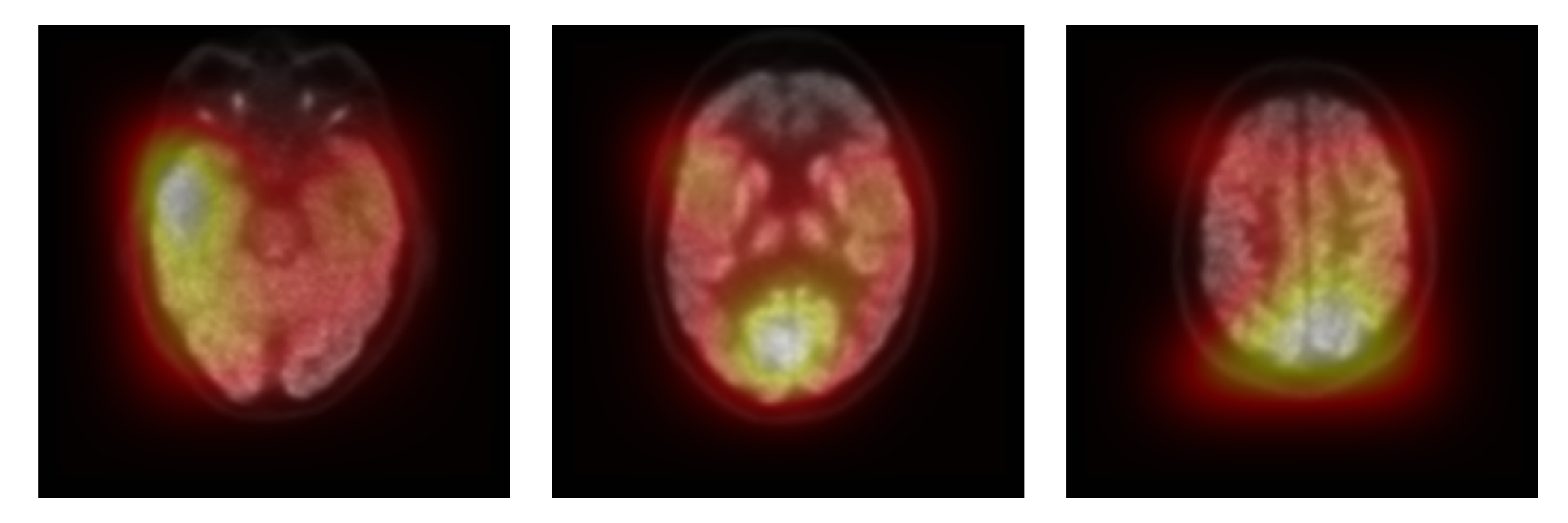}
    \\[-1ex]
    \subfloat[Cases with attention maps primarily overlapping with brain regions]{\includegraphics[width=.47\textwidth]{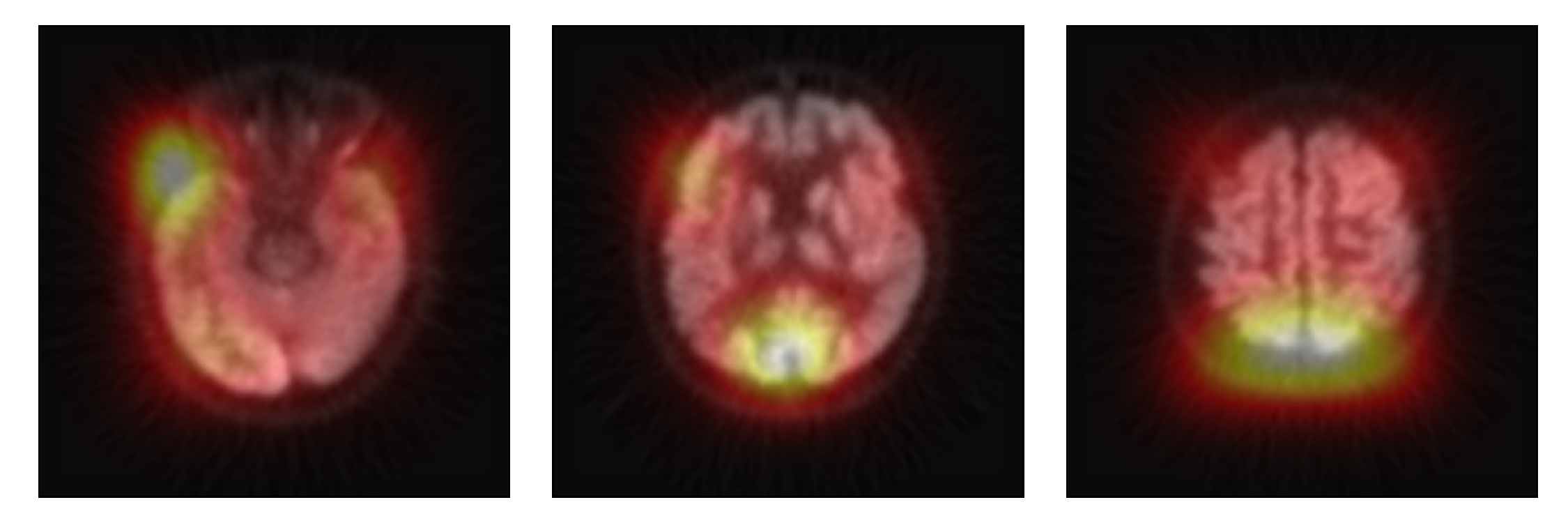}}
    \\
    \includegraphics[width=.47\textwidth]{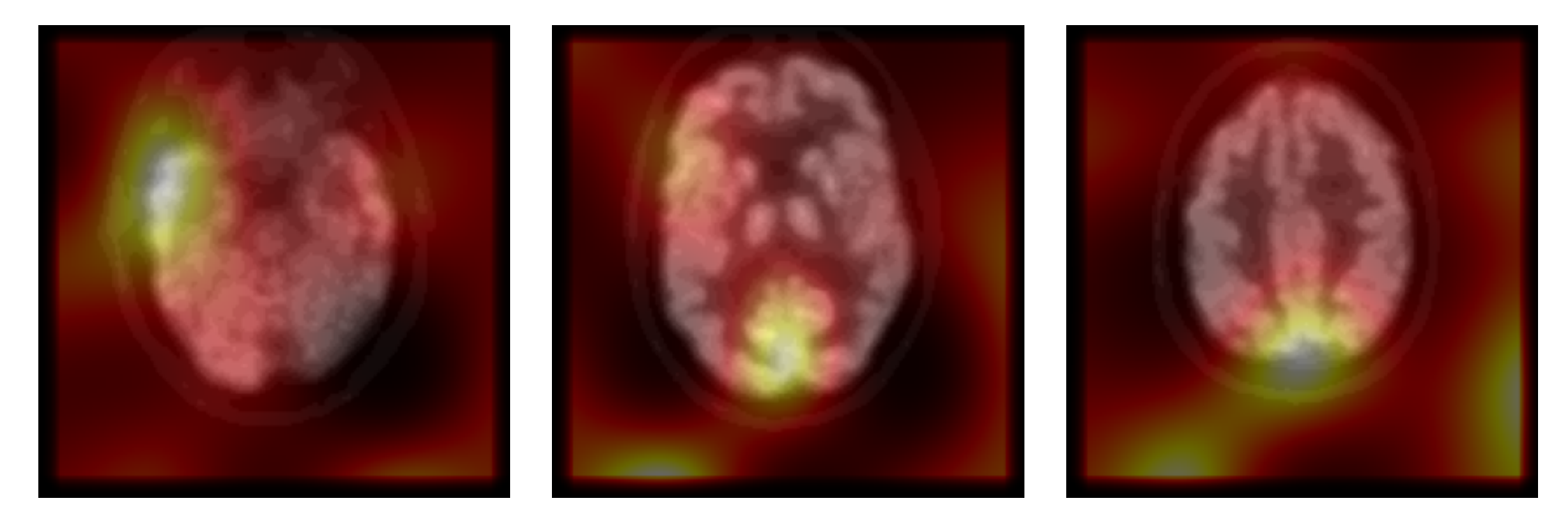}    
    \\[-1ex]
    \subfloat[Cases with attention maps overlapping with both brain and background regions]{\includegraphics[width=.47\textwidth]{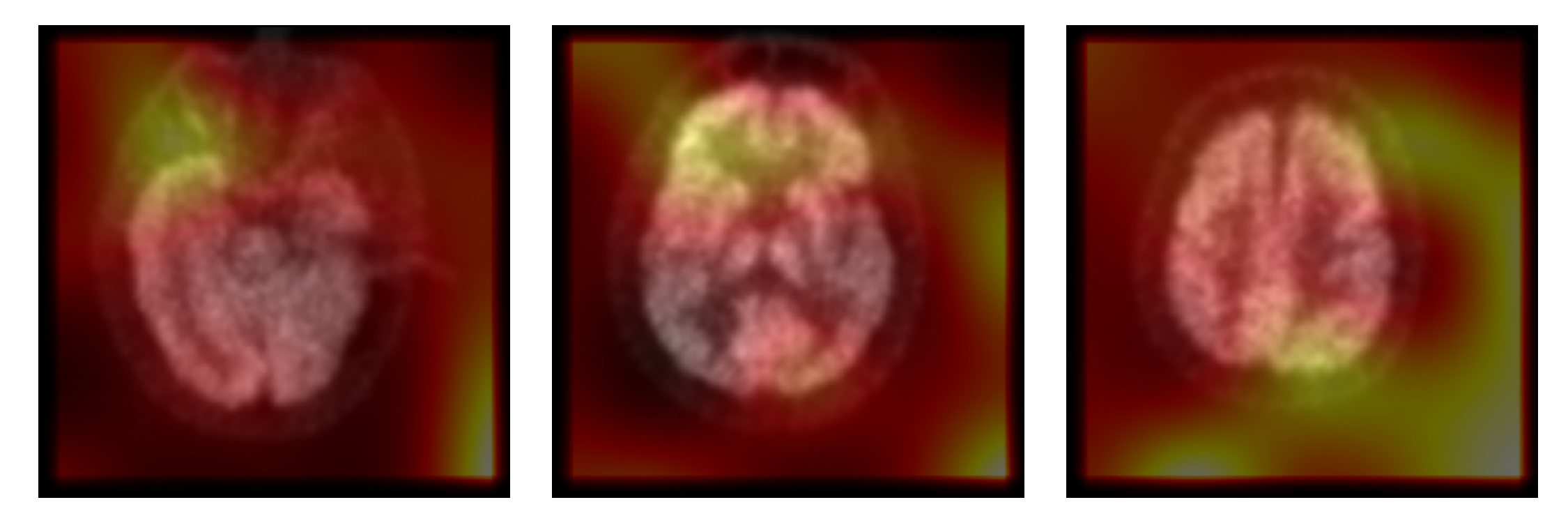}}
    \caption{Attention maps on axial FDG-PET slices. Three axial slices on each row belong to the same PET scan.}
    \label{fig:adni_attn_examples}
\end{figure}

\linelabel{ln:attn_limitation}\update{The second limitation of the present study, is that attention maps produced by transformers act as human-friendly signals of our model, and should be carefully used in practice with expert knowledge in the domain. Transformers may highlight areas not associated with the body part, which can be seen in~\cref{fig:adni_attn_examples} as well as in other studies~\cite{petit2021u,odusami2022intelligent,rao2021counterfactual}.}
% Second, we treat our work as a disease forecasting framework that can produce self-attention-based heatmaps for each prediction; however, we did not deep dive into the clinical relevance of highlighted areas or clinical variables. Therefore, those heatmaps should be considered as non-clinical signals from our models, and any clinical interpretations should be carefully made based on them.}

\update{Lastly, we primarily utilized the transformer proposed by~\cite{dosovitskiy2020image}. More efficient and advanced transformers such as~\cite{kitaev2020reformer,zhou2021informer,chen2021autoformer,yang2021causal} could be further investigated to integrate into the framework.}

\linelabel{ln:summary}To conclude, to our knowledge, this is not only the first study in the realm of OA, but also the first work on AD clinical status prognosis prediction from the multi-modal setup that includes raw 3D scans and scalar variables. The developed method can be of interest to other fields, where forecasting of calibrated disease trajectory is of interest. 
\linelabel{ln:code}\updatetwo{An implementation of our method is made publicly available at \url{https://github.com/Oulu-IMEDS/CLIMATv2}.}

% \todo[inline]{The discussion about the heatmaps is strange. I cannot follow the logic. Needs work.}

% \iffalse
\section{Acknowledgement}
\label{sc:acknowledgement}
\update{
The OAI is a public-private partnership comprised of five contracts (N01- AR-2-2258; N01-AR-2-2259; N01-AR-2- 2260; N01-AR-2-2261; N01-AR-2-2262) funded by the National Institutes of Health, a branch of the Department of Health and Human Services, and conducted by the OAI Study Investigators. Private funding partners include Merck Research Laboratories; Novartis Pharmaceuticals Corporation, GlaxoSmithKline; and Pfizer, Inc. Private sector funding for the OAI is managed by the Foundation for the National Institutes of Health.} 

\update{Data collection and sharing for this project was funded by the Alzheimer's Disease Neuroimaging Initiative (ADNI) (National Institutes of Health Grant U01 AG024904) and DOD ADNI (Department of Defense award number W81XWH-12-2-0012). ADNI is funded by the National Institute on Aging, the National Institute of Biomedical Imaging and Bioengineering, and through generous contributions from the following: AbbVie, Alzheimer's Association; Alzheimer's Drug Discovery Foundation; Araclon Biotech; BioClinica, Inc.; Biogen; Bristol-Myers Squibb Company; CereSpir, Inc.; Cogstate; Eisai Inc.; Elan Pharmaceuticals, Inc.; Eli Lilly and Company; EuroImmun; F. Hoffmann-La Roche Ltd and its affiliated company Genentech, Inc.; Fujirebio; GE Healthcare; IXICO Ltd.; Janssen Alzheimer Immunotherapy Research \& Development, LLC.; Johnson \& Johnson Pharmaceutical Research \& Development LLC.; Lumosity; Lundbeck; Merck \& Co., Inc.; Meso Scale Diagnostics, LLC.; NeuroRx Research; Neurotrack Technologies; Novartis Pharmaceuticals Corporation; Pfizer Inc.; Piramal Imaging; Servier; Takeda Pharmaceutical Company; and Transition Therapeutics. The Canadian Institutes of Health Research is providing funds to support ADNI clinical sites in Canada. Private sector contributions are facilitated by the Foundation for the National Institutes of Health (\url{www.fnih.org}). The grantee organization is the Northern California Institute for Research and Education, and the study is coordinated by the Alzheimer's Therapeutic Research Institute at the University of Southern California. ADNI data are disseminated by the Laboratory for Neuro Imaging at the University of Southern California.}

\update{The authors wish to acknowledge CSC – IT Center for Science, Finland, for generous computational resources.}

\update{We would like to acknowledge the strategic funding of the University of Oulu, Infotech Oulu Focus Institute, and Sigrid Juselius Foundation, Finland. The publication was also supported by funding from the
Academy of Finland (Profi6 336449 funding program).} \linelabel{ln:acknowledgement}\updatetwo{We also acknowledge funding from the Alfred Kordelinin foundation (220211).}

\update{We acknowledge funding from the Flemish Government under the “Onderzoeksprogramma Artificiële Intelligentie (AI) Vlaanderen" programme.}

\update{Dr. Claudia Lindner is acknowledged for providing BoneFinder. Phuoc Dat Nguyen is acknowledged for discussions about transformers.}

% \fi

\bibliography{bibtex/IEEEabrv.bib,bibtex/reference.bib}{}

\begin{thebibliography}{10}
\providecommand{\url}[1]{#1}
\csname url@samestyle\endcsname
\providecommand{\newblock}{\relax}
\providecommand{\bibinfo}[2]{#2}
\providecommand{\BIBentrySTDinterwordspacing}{\spaceskip=0pt\relax}
\providecommand{\BIBentryALTinterwordstretchfactor}{4}
\providecommand{\BIBentryALTinterwordspacing}{\spaceskip=\fontdimen2\font plus
\BIBentryALTinterwordstretchfactor\fontdimen3\font minus
  \fontdimen4\font\relax}
\providecommand{\BIBforeignlanguage}[2]{{%
\expandafter\ifx\csname l@#1\endcsname\relax
\typeout{** WARNING: IEEEtran.bst: No hyphenation pattern has been}%
\typeout{** loaded for the language `#1'. Using the pattern for}%
\typeout{** the default language instead.}%
\else
\language=\csname l@#1\endcsname
\fi
#2}}
\providecommand{\BIBdecl}{\relax}
\BIBdecl

\bibitem{desai2020comparison}
R.~J. Desai, S.~V. Wang, M.~Vaduganathan, T.~Evers, and S.~Schneeweiss,
  ``Comparison of machine learning methods with traditional models for use of
  administrative claims with electronic medical records to predict heart
  failure outcomes,'' \emph{JAMA network open}, vol.~3, no.~1, pp.
  e1\,918\,962--e1\,918\,962, 2020.

\bibitem{boutet2021predicting}
A.~Boutet, R.~Madhavan, G.~J. Elias, S.~E. Joel, R.~Gramer, M.~Ranjan,
  V.~Paramanandam, D.~Xu, J.~Germann, A.~Loh \emph{et~al.}, ``Predicting
  optimal deep brain stimulation parameters for parkinson’s disease using
  functional mri and machine learning,'' \emph{Nature communications}, vol.~12,
  no.~1, pp. 1--13, 2021.

\bibitem{cheerla2019deep}
A.~Cheerla and O.~Gevaert, ``Deep learning with multimodal representation for
  pancancer prognosis prediction,'' \emph{Bioinformatics}, vol.~35, no.~14, pp.
  i446--i454, 2019.

\bibitem{tran2021deep}
K.~A. Tran, O.~Kondrashova, A.~Bradley, E.~D. Williams, J.~V. Pearson, and
  N.~Waddell, ``Deep learning in cancer diagnosis, prognosis and treatment
  selection,'' \emph{Genome Medicine}, vol.~13, no.~1, pp. 1--17, 2021.

\bibitem{milanez2020cancer}
P.~Milanez-Almeida, A.~J. Martins, R.~N. Germain, and J.~S. Tsang, ``Cancer
  prognosis with shallow tumor rna sequencing,'' \emph{Nature Medicine},
  vol.~26, no.~2, pp. 188--192, 2020.

\bibitem{mei2020artificial}
X.~Mei, H.-C. Lee, K.-y. Diao, M.~Huang, B.~Lin, C.~Liu, Z.~Xie, Y.~Ma, P.~M.
  Robson, M.~Chung \emph{et~al.}, ``Artificial intelligence--enabled rapid
  diagnosis of patients with covid-19,'' \emph{Nature medicine}, vol.~26,
  no.~8, pp. 1224--1228, 2020.

\bibitem{glyn2015osteoarthritis}
S.~Glyn-Jones, A.~Palmer, R.~Agricola, A.~Price, T.~Vincent, H.~Weinans, and
  A.~Carr, ``Osteoarthritis,'' \emph{The Lancet}, vol. 386, no. 9991, pp.
  376--387, 2015.

\bibitem{AD2021}
``2021 alzheimer's disease facts and figures,'' \emph{Alzheimer's \& Dementia},
  vol.~17, no.~3, pp. 327--406, 2021.

\bibitem{heidari2011knee}
B.~Heidari, ``Knee osteoarthritis prevalence, risk factors, pathogenesis and
  features: Part i,'' \emph{Caspian journal of internal medicine}, vol.~2,
  no.~2, p. 205, 2011.

\bibitem{lee2021imaging}
L.~S. Lee, P.~K. Chan, W.~C. Fung, V.~W.~K. Chan, C.~H. Yan, and K.~Y. Chiu,
  ``Imaging of knee osteoarthritis: A review of current evidence and clinical
  guidelines,'' \emph{Musculoskeletal Care}, vol.~19, no.~3, pp. 363--374,
  2021.

\bibitem{kellgren1957radiological}
J.~Kellgren and J.~Lawrence, ``Radiological assessment of osteo-arthrosis,''
  \emph{Annals of the rheumatic diseases}, vol.~16, no.~4, p. 494, 1957.

\bibitem{altman2007atlas}
R.~D. Altman and G.~Gold, ``Atlas of individual radiographic features in
  osteoarthritis, revised,'' \emph{Osteoarthritis and cartilage}, vol.~15, pp.
  A1--A56, 2007.

\bibitem{rezucs2021pathogenesis}
E.~Rezu{\c{s}}, A.~Burlui, A.~Cardoneanu, L.~A. Macovei, B.~I. Tamba, and
  C.~Rezu{\c{s}}, ``From pathogenesis to therapy in knee osteoarthritis:
  bench-to-bedside,'' \emph{International Journal of Molecular Sciences},
  vol.~22, no.~5, p. 2697, 2021.

\bibitem{rasmussen2019alzheimer}
J.~Rasmussen and H.~Langerman, ``Alzheimer’s disease--why we need early
  diagnosis,'' \emph{Degenerative neurological and neuromuscular disease},
  vol.~9, p. 123, 2019.

\bibitem{tiulpin2019multimodal}
A.~Tiulpin, S.~Klein, S.~M. Bierma-Zeinstra, J.~Thevenot, E.~Rahtu, J.~van
  Meurs, E.~H. Oei, and S.~Saarakkala, ``Multimodal machine learning-based knee
  osteoarthritis progression prediction from plain radiographs and clinical
  data,'' \emph{Scientific reports}, vol.~9, no.~1, pp. 1--11, 2019.

\bibitem{widera2020multi}
P.~Widera, P.~M. Welsing, C.~Ladel, J.~Loughlin, F.~P. Lafeber, F.~P. Dop,
  J.~Larkin, H.~Weinans, A.~Mobasheri, and J.~Bacardit, ``Multi-classifier
  prediction of knee osteoarthritis progression from incomplete imbalanced
  longitudinal data,'' \emph{Scientific Reports}, vol.~10, no.~1, pp. 1--15,
  2020.

\bibitem{guan2020deep}
B.~Guan, F.~Liu, A.~Haj-Mirzaian, S.~Demehri, A.~Samsonov, T.~Neogi,
  A.~Guermazi, and R.~Kijowski, ``Deep learning risk assessment models for
  predicting progression of radiographic medial joint space loss over a
  48-month follow-up period,'' \emph{Osteoarthritis and cartilage}, vol.~28,
  no.~4, pp. 428--437, 2020.

\bibitem{leung2020prediction}
K.~Leung, B.~Zhang, J.~Tan, Y.~Shen, K.~J. Geras, J.~S. Babb, K.~Cho, G.~Chang,
  and C.~M. Deniz, ``Prediction of total knee replacement and diagnosis of
  osteoarthritis by using deep learning on knee radiographs: data from the
  osteoarthritis initiative,'' \emph{Radiology}, vol. 296, no.~3, pp. 584--593,
  2020.

\bibitem{tolpadi2020deep}
A.~A. Tolpadi, J.~J. Lee, V.~Pedoia, and S.~Majumdar, ``Deep learning predicts
  total knee replacement from magnetic resonance images,'' \emph{Scientific
  reports}, vol.~10, no.~1, pp. 1--12, 2020.

\bibitem{jung2019unified}
W.~Jung, A.~W. Mulyadi, and H.-I. Suk, ``Unified modeling of imputation,
  forecasting, and prediction for ad progression,'' in \emph{International
  Conference on Medical Image Computing and Computer-Assisted
  Intervention}.\hskip 1em plus 0.5em minus 0.4em\relax Springer, 2019, pp.
  168--176.

\bibitem{lu2018multimodal}
D.~Lu, K.~Popuri, G.~W. Ding, R.~Balachandar, and M.~F. Beg, ``Multimodal and
  multiscale deep neural networks for the early diagnosis of alzheimer’s
  disease using structural mr and fdg-pet images,'' \emph{Scientific reports},
  vol.~8, no.~1, pp. 1--13, 2018.

\bibitem{nguyen2022climat}
H.~H. Nguyen, S.~Saarakkala, M.~B. Blaschko, and A.~Tiulpin, ``Climat:
  Clinically-inspired multi-agent transformers for knee osteoarthritis
  trajectory forecasting,'' in \emph{2022 IEEE 19th International Symposium on
  Biomedical Imaging (ISBI)}.\hskip 1em plus 0.5em minus 0.4em\relax IEEE,
  2022, pp. 1--5.

\bibitem{ghazi2019training}
M.~M. Ghazi, M.~Nielsen, A.~Pai, M.~J. Cardoso, M.~Modat, S.~Ourselin,
  L.~S{\o}rensen, A.~D.~N. Initiative \emph{et~al.}, ``Training recurrent
  neural networks robust to incomplete data: application to alzheimer’s
  disease progression modeling,'' \emph{Medical image analysis}, vol.~53, pp.
  39--46, 2019.

\bibitem{albright2019forecasting}
J.~Albright, A.~D.~N. Initiative \emph{et~al.}, ``Forecasting the progression
  of alzheimer's disease using neural networks and a novel preprocessing
  algorithm,'' \emph{Alzheimer's \& Dementia: Translational Research \&
  Clinical Interventions}, vol.~5, pp. 483--491, 2019.

\bibitem{vaswani2017attention}
A.~Vaswani, N.~Shazeer, N.~Parmar, J.~Uszkoreit, L.~Jones, A.~N. Gomez,
  {\L}.~Kaiser, and I.~Polosukhin, ``Attention is all you need,'' in
  \emph{Advances in neural information processing systems}, 2017, pp.
  5998--6008.

\bibitem{devlin2018bert}
J.~Devlin, M.-W. Chang, K.~Lee, and K.~Toutanova, ``Bert: Pre-training of deep
  bidirectional transformers for language understanding,'' \emph{arXiv preprint
  arXiv:1810.04805}, 2018.

\bibitem{dosovitskiy2020image}
A.~Dosovitskiy, L.~Beyer, A.~Kolesnikov, D.~Weissenborn, X.~Zhai,
  T.~Unterthiner, M.~Dehghani, M.~Minderer, G.~Heigold, S.~Gelly \emph{et~al.},
  ``An image is worth 16x16 words: Transformers for image recognition at
  scale,'' \emph{arXiv preprint arXiv:2010.11929}, vol.~1, 2020.

\bibitem{girdhar2019video}
R.~Girdhar, J.~Carreira, C.~Doersch, and A.~Zisserman, ``Video action
  transformer network,'' in \emph{Proceedings of the IEEE/CVF Conference on
  Computer Vision and Pattern Recognition}, 2019, pp. 244--253.

\bibitem{arnab2021vivit}
A.~Arnab, M.~Dehghani, G.~Heigold, C.~Sun, M.~Lu{\v{c}}i{\'c}, and C.~Schmid,
  ``Vivit: A video vision transformer,'' \emph{arXiv preprint
  arXiv:2103.15691}, 2021.

\bibitem{hassani2021escaping}
A.~Hassani, S.~Walton, N.~Shah, A.~Abuduweili, J.~Li, and H.~Shi, ``Escaping
  the big data paradigm with compact transformers,'' \emph{arXiv preprint
  arXiv:2104.05704}, 2021.

\bibitem{hu2021transformer}
S.~Hu, E.~Fridgeirsson, G.~van Wingen, and M.~Welling, ``Transformer-based deep
  survival analysis,'' in \emph{Survival Prediction-Algorithms, Challenges and
  Applications}.\hskip 1em plus 0.5em minus 0.4em\relax PMLR, 2021, pp.
  132--148.

\bibitem{radford2021learning}
A.~Radford, J.~W. Kim, C.~Hallacy, A.~Ramesh, G.~Goh, S.~Agarwal, G.~Sastry,
  A.~Askell, P.~Mishkin, J.~Clark \emph{et~al.}, ``Learning transferable visual
  models from natural language supervision,'' in \emph{International Conference
  on Machine Learning}.\hskip 1em plus 0.5em minus 0.4em\relax PMLR, 2021, pp.
  8748--8763.

\bibitem{jans2013optimizing}
L.~Jans, J.~Bosmans, K.~Verstraete, and R.~Achten, ``Optimizing communication
  between the radiologist and the general practitioner,'' \emph{JBR-BTR},
  vol.~96, no.~6, pp. 388--390, 2013.

\bibitem{liu2023joint}
L.~Liu, J.~Chang, P.~Zhang, Q.~Ma, H.~Zhang, T.~Sun, and H.~Qiao, ``A joint
  multi-modal learning method for early-stage knee osteoarthritis disease
  classification,'' \emph{Heliyon}, vol.~9, no.~4, 2023.

\bibitem{guo2017calibration}
C.~Guo, G.~Pleiss, Y.~Sun, and K.~Q. Weinberger, ``On calibration of modern
  neural networks,'' in \emph{International conference on machine
  learning}.\hskip 1em plus 0.5em minus 0.4em\relax PMLR, 2017, pp. 1321--1330.

\bibitem{ba2016layer}
J.~L. Ba, J.~R. Kiros, and G.~E. Hinton, ``Layer normalization,'' \emph{arXiv
  preprint arXiv:1607.06450}, 2016.

\bibitem{hendrycks2020gaussian}
D.~Hendrycks and K.~Gimpel, ``Gaussian error linear units ({GELUs}),'' 2020,
  arXiv:1606.08415.

\bibitem{chu2021conditional}
X.~Chu, Z.~Tian, B.~Zhang, X.~Wang, X.~Wei, H.~Xia, and C.~Shen, ``Conditional
  positional encodings for vision transformers,'' \emph{arXiv preprint
  arXiv:2102.10882}, 2021.

\bibitem{pan2021scalable}
Z.~Pan, B.~Zhuang, J.~Liu, H.~He, and J.~Cai, ``Scalable vision transformers
  with hierarchical pooling,'' in \emph{Proceedings of the IEEE/cvf
  international conference on computer vision}, 2021, pp. 377--386.

\bibitem{park2022vision}
N.~Park and S.~Kim, ``How do vision transformers work?'' \emph{arXiv preprint
  arXiv:2202.06709}, 2022.

\bibitem{kendall2018multi}
A.~Kendall, Y.~Gal, and R.~Cipolla, ``Multi-task learning using uncertainty to
  weigh losses for scene geometry and semantics,'' in \emph{Proceedings of the
  IEEE conference on computer vision and pattern recognition}, 2018, pp.
  7482--7491.

\bibitem{tiulpin2020automatic}
A.~Tiulpin and S.~Saarakkala, ``Automatic grading of individual knee
  osteoarthritis features in plain radiographs using deep convolutional neural
  networks,'' \emph{Diagnostics}, vol.~10, no.~11, p. 932, 2020.

\bibitem{lindner2013fully}
C.~Lindner, S.~Thiagarajah, J.~M. Wilkinson, G.~A. Wallis, T.~F. Cootes,
  arcOGEN Consortium \emph{et~al.}, ``Fully automatic segmentation of the
  proximal femur using random forest regression voting,'' \emph{IEEE
  transactions on medical imaging}, vol.~32, no.~8, pp. 1462--1472, 2013.

\bibitem{brett2020nipy}
M.~Brett, C.~J. Markiewicz, M.~Hanke, M.-A. C{\^o}t{\'e}, B.~Cipollini,
  P.~McCarthy, D.~Jarecka, C.~Cheng, Y.~Halchenko, M.~Cottaar \emph{et~al.},
  ``nipy/nibabel: 3.2. 1,'' \emph{Zenodo}, 2020.

\bibitem{paszke2019pytorch}
A.~Paszke, S.~Gross, F.~Massa, A.~Lerer, J.~Bradbury, G.~Chanan, T.~Killeen,
  Z.~Lin, N.~Gimelshein, L.~Antiga \emph{et~al.}, ``Pytorch: An imperative
  style, high-performance deep learning library,'' \emph{arXiv preprint
  arXiv:1912.01703}, 2019.

\bibitem{kingma2014adam}
D.~P. Kingma and J.~Ba, ``Adam: A method for stochastic optimization,''
  \emph{arXiv preprint arXiv:1412.6980}, 2014.

\bibitem{he2016deep}
K.~He, X.~Zhang, S.~Ren, and J.~Sun, ``Deep residual learning for image
  recognition,'' in \emph{Proceedings of the IEEE conference on computer vision
  and pattern recognition}, 2016, pp. 770--778.

\bibitem{deng2009imagenet}
J.~Deng, W.~Dong, R.~Socher, L.-J. Li, K.~Li, and L.~Fei-Fei, ``Imagenet: A
  large-scale hierarchical image database,'' in \emph{2009 IEEE conference on
  computer vision and pattern recognition}.\hskip 1em plus 0.5em minus
  0.4em\relax Ieee, 2009, pp. 248--255.

\bibitem{kopuklu2019resource}
O.~Kopuklu, N.~Kose, A.~Gunduz, and G.~Rigoll, ``Resource efficient 3d
  convolutional neural networks,'' in \emph{Proceedings of the IEEE/CVF
  International Conference on Computer Vision Workshops}, 2019, pp. 0--0.

\bibitem{carreira2017quo}
J.~Carreira and A.~Zisserman, ``Quo vadis, action recognition? a new model and
  the kinetics dataset,'' in \emph{proceedings of the IEEE Conference on
  Computer Vision and Pattern Recognition}, 2017, pp. 6299--6308.

\bibitem{cho2014properties}
K.~Cho, B.~Van~Merri{\"e}nboer, D.~Bahdanau, and Y.~Bengio, ``On the properties
  of neural machine translation: Encoder-decoder approaches,'' \emph{arXiv
  preprint arXiv:1409.1259}, 2014.

\bibitem{hochreiter1997long}
S.~Hochreiter and J.~Schmidhuber, ``Long short-term memory,'' \emph{Neural
  computation}, vol.~9, no.~8, pp. 1735--1780, 1997.

\bibitem{kitaev2020reformer}
N.~Kitaev, {\L}.~Kaiser, and A.~Levskaya, ``Reformer: The efficient
  transformer,'' \emph{arXiv preprint arXiv:2001.04451}, 2020.

\bibitem{zhou2021informer}
H.~Zhou, S.~Zhang, J.~Peng, S.~Zhang, J.~Li, H.~Xiong, and W.~Zhang,
  ``Informer: Beyond efficient transformer for long sequence time-series
  forecasting,'' in \emph{Proceedings of the AAAI conference on artificial
  intelligence}, vol.~35, no.~12, 2021, pp. 11\,106--11\,115.

\bibitem{chen2021autoformer}
M.~Chen, H.~Peng, J.~Fu, and H.~Ling, ``Autoformer: Searching transformers for
  visual recognition,'' in \emph{Proceedings of the IEEE/CVF international
  conference on computer vision}, 2021, pp. 12\,270--12\,280.

\bibitem{brodersen2010balanced}
K.~H. Brodersen, C.~S. Ong, K.~E. Stephan, and J.~M. Buhmann, ``The balanced
  accuracy and its posterior distribution,'' in \emph{2010 20th international
  conference on pattern recognition}.\hskip 1em plus 0.5em minus 0.4em\relax
  IEEE, 2010, pp. 3121--3124.

\bibitem{hand2001simple}
D.~J. Hand and R.~J. Till, ``A simple generalisation of the area under the roc
  curve for multiple class classification problems,'' \emph{Machine learning},
  vol.~45, no.~2, pp. 171--186, 2001.

\bibitem{naeini2015obtaining}
M.~P. Naeini, G.~Cooper, and M.~Hauskrecht, ``Obtaining well calibrated
  probabilities using bayesian binning,'' in \emph{Twenty-Ninth AAAI Conference
  on Artificial Intelligence}, 2015.

\bibitem{wilcoxon1992individual}
F.~Wilcoxon, ``Individual comparisons by ranking methods,'' in
  \emph{Breakthroughs in statistics}.\hskip 1em plus 0.5em minus 0.4em\relax
  Springer, 1992, pp. 196--202.

\bibitem{dunn1961multiple}
O.~J. Dunn, ``Multiple comparisons among means,'' \emph{Journal of the American
  statistical association}, vol.~56, no. 293, pp. 52--64, 1961.

\bibitem{lin2017focal}
T.-Y. Lin, P.~Goyal, R.~Girshick, K.~He, and P.~Doll{\'a}r, ``Focal loss for
  dense object detection,'' in \emph{Proceedings of the IEEE international
  conference on computer vision}, 2017, pp. 2980--2988.

\bibitem{mukhoti2020calibrating}
J.~Mukhoti, V.~Kulharia, A.~Sanyal, S.~Golodetz, P.~H. Torr, and P.~K. Dokania,
  ``Calibrating deep neural networks using focal loss,'' \emph{arXiv preprint
  arXiv:2002.09437}, 2020.

\bibitem{leon2005intercondylar}
H.~O. Le{\'o}n, C.~E.~R. Blanco, T.~B. Guthrie, and O.~J.~N. Mart{\'\i}nez,
  ``Intercondylar notch stenosis in degenerative arthritis of the knee,''
  \emph{Arthroscopy: The Journal of Arthroscopic \& Related Surgery}, vol.~21,
  no.~3, pp. 294--302, 2005.

\bibitem{hallam2020neural}
B.~Hallam, J.~Chan, S.~G. Costafreda, R.~Bhome, and J.~Huntley, ``What are the
  neural correlates of meta-cognition and anosognosia in alzheimer's disease? a
  systematic review,'' \emph{Neurobiology of aging}, vol.~94, pp. 250--264,
  2020.

\bibitem{bird2005genetic}
T.~D. Bird, ``Genetic factors in alzheimer's disease,'' \emph{New England
  Journal of Medicine}, vol. 352, no.~9, pp. 862--864, 2005.

\bibitem{li2022validation}
Y.~Li, S.~E. Schindler, J.~G. Bollinger, V.~Ovod, K.~G. Mawuenyega, M.~W.
  Weiner, L.~M. Shaw, C.~L. Masters, C.~J. Fowler, J.~Q. Trojanowski
  \emph{et~al.}, ``Validation of plasma amyloid-$\beta$ 42/40 for detecting
  alzheimer disease amyloid plaques,'' \emph{Neurology}, vol.~98, no.~7, pp.
  e688--e699, 2022.

\bibitem{elsken2019neural}
T.~Elsken, J.~H. Metzen, and F.~Hutter, ``Neural architecture search: A
  survey,'' \emph{The Journal of Machine Learning Research}, vol.~20, no.~1,
  pp. 1997--2017, 2019.

\bibitem{petit2021u}
O.~Petit, N.~Thome, C.~Rambour, L.~Themyr, T.~Collins, and L.~Soler, ``U-net
  transformer: Self and cross attention for medical image segmentation,'' in
  \emph{International Workshop on Machine Learning in Medical Imaging}.\hskip
  1em plus 0.5em minus 0.4em\relax Springer, 2021, pp. 267--276.

\bibitem{odusami2022intelligent}
M.~Odusami, R.~Maskeli{\=u}nas, and R.~Dama{\v{s}}evi{\v{c}}ius, ``An
  intelligent system for early recognition of alzheimer’s disease using
  neuroimaging,'' \emph{Sensors}, vol.~22, no.~3, p. 740, 2022.

\bibitem{rao2021counterfactual}
Y.~Rao, G.~Chen, J.~Lu, and J.~Zhou, ``Counterfactual attention learning for
  fine-grained visual categorization and re-identification,'' in
  \emph{Proceedings of the IEEE/CVF International Conference on Computer
  Vision}, 2021, pp. 1025--1034.

\bibitem{yang2021causal}
X.~Yang, H.~Zhang, G.~Qi, and J.~Cai, ``Causal attention for vision-language
  tasks,'' in \emph{Proceedings of the IEEE/CVF conference on computer vision
  and pattern recognition}, 2021, pp. 9847--9857.

\end{thebibliography}
\bibliographystyle{IEEEtran}

\end{document}